\documentclass[aps,twocolumn,amsmath,amssymb,superscriptaddress]{revtex4}
\usepackage{graphicx}
\usepackage{dcolumn}
\usepackage{bm}
\usepackage{color}
\begin{document}
\title{Thermoelectric generation based on spin Seebeck effects}
\author{Ken-ichi Uchida}
\email{kuchida@imr.tohoku.ac.jp}
\affiliation{Institute for Materials Research, Tohoku University, Sendai 980-8577, Japan}
\affiliation{PRESTO, Japan Science and Technology Agency, Saitama 332-0012, Japan}
\author{Hiroto Adachi}
\affiliation{Advanced Science Research Center, Japan Atomic Energy Agency, Tokai 319-1195, Japan}
\affiliation{Spin Quantum Rectification Project, ERATO, Japan Science and Technology Agency, Sendai 980-8577, Japan}
\author{Takashi Kikkawa}
\affiliation{Institute for Materials Research, Tohoku University, Sendai 980-8577, Japan}
\affiliation{WPI Advanced Institute for Materials Research, Tohoku University, Sendai 980-8577, Japan}
\author{Akihiro Kirihara}
\affiliation{Spin Quantum Rectification Project, ERATO, Japan Science and Technology Agency, Sendai 980-8577, Japan}
\affiliation{Smart Energy Research Laboratories, NEC Corporation, Tsukuba 305-8501, Japan}
\author{Masahiko Ishida}
\affiliation{Spin Quantum Rectification Project, ERATO, Japan Science and Technology Agency, Sendai 980-8577, Japan}
\affiliation{Smart Energy Research Laboratories, NEC Corporation, Tsukuba 305-8501, Japan}
\author{Shinichi Yorozu}
\affiliation{Spin Quantum Rectification Project, ERATO, Japan Science and Technology Agency, Sendai 980-8577, Japan}
\affiliation{Smart Energy Research Laboratories, NEC Corporation, Tsukuba 305-8501, Japan}
\author{Sadamichi Maekawa}
\affiliation{Advanced Science Research Center, Japan Atomic Energy Agency, Tokai 319-1195, Japan}
\affiliation{Spin Quantum Rectification Project, ERATO, Japan Science and Technology Agency, Sendai 980-8577, Japan}
\author{Eiji Saitoh}
\affiliation{Institute for Materials Research, Tohoku University, Sendai 980-8577, Japan}
\affiliation{Advanced Science Research Center, Japan Atomic Energy Agency, Tokai 319-1195, Japan}
\affiliation{Spin Quantum Rectification Project, ERATO, Japan Science and Technology Agency, Sendai 980-8577, Japan}
\affiliation{WPI Advanced Institute for Materials Research, Tohoku University, Sendai 980-8577, Japan}
\date{\today}
\begin{abstract}
The spin Seebeck effect (SSE) refers to the generation of a spin current as a result of a temperature gradient in magnetic materials including insulators. The SSE is applicable to thermoelectric generation because the thermally generated spin current can be converted into a charge current via spin-orbit interaction in conductive materials adjacent to the magnets. The insulator-based SSE device exhibits unconventional characteristics potentially useful for thermoelectric applications, such as simple structure, device-design flexibility, and convenient scaling capability. In this article, we review recent studies on the SSE from the viewpoint of thermoelectric applications. Firstly, we introduce the thermoelectric generation process and measurement configuration of the SSE, followed by showing fundamental characteristics of the SSE device. Secondly, a theory of the thermoelectric conversion efficiency of the SSE device is presented, which clarifies the difference between the SSE and conventional thermoelectric effects and the efficiency limit of the SSE device. Finally, we show preliminary demonstrations of the SSE in various device structures for future thermoelectric applications and discuss prospects of the SSE-based thermoelectric technologies. \par
KEYWORDS | spintronics; spin current; spin Seebeck effect; inverse spin Hall effect; anomalous Nernst effect; thermoelectric generation; magnetic material; thin film
\end{abstract}
\maketitle
%
%------------main-text----------------------------------
%
%%%%%%%%%%%%%%%%%%%%%%%%%%%%%%%%%%%%%%%%%%%%%%%%%%%%
\section{INTRODUCTION} \label{sec:intro}
%%%%%%%%%%%%%%%%%%%%%%%%%%%%%%%%%%%%%%%%%%%%%%%%%%%%
%
%
Heat is an omnipresent and abundant source of energy, and thus the efficient utilization of thermal energy, such as waste heat recovery and solar heat power generation, is indispensable for the realization of future sustainable society. Thermoelectric generation is one of the promising technologies for making effective use of heat since it enables direct conversion from thermal energy to electrical power \cite{DiSalvo1999Science,Bell2008Science,Sundarraj2014RSCAdv}. As thermoelectric devices consist of solid-state materials and have no moving parts, they are silent, reliable, and scalable. Most of the prevalent thermoelectric generation technologies are based on the Seebeck effect, discovered by T. J. Seebeck in 1821 \cite{Rowe,Goldsmid}. Although a variety of thermoelectric effects were found to appear in electric conductors subjected to temperature gradients after the discovery of the Seebeck effect \cite{spincaloritronics-Heremans}, the earliest thermoelectric phenomenon is still the key player in the thermoelectric technology because of its relatively-high output power and versatility. \par
\begin{figure*}[tb]
\begin{center}
\includegraphics{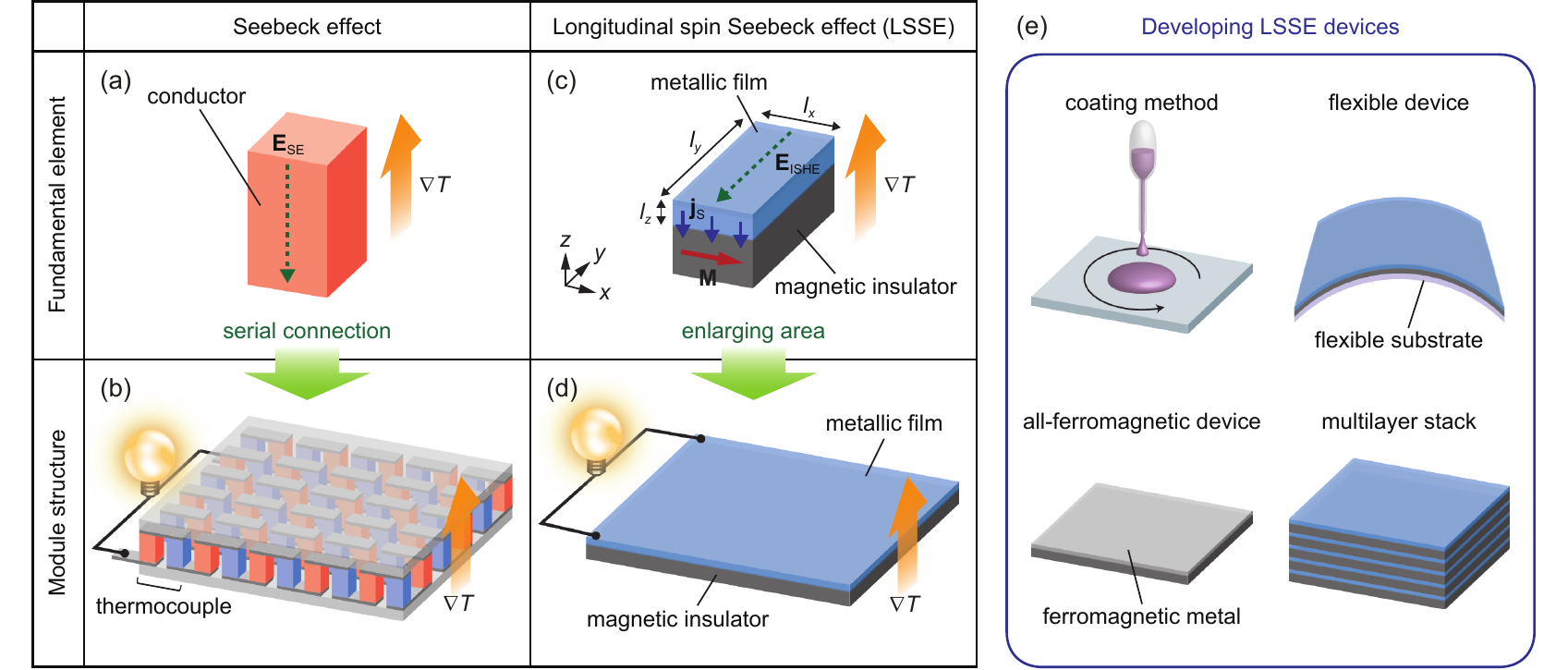}
\caption{(a),(b) Schematic illustrations of the fundamental element and module structure of the thermoelectric device based on the Seebeck effect. (c),(d) Schematic illustrations of the fundamental element and module structure of the thermoelectric device based on the longitudinal spin Seebeck effect (LSSE). $\nabla T$, ${\bf E}_{\rm SE(ISHE)}$, ${\bf M}$, and ${\bf j}_{\rm S}$ denote the temperature gradient, electric field generated by the Seebeck effect [inverse spin Hall effect (ISHE)], magnetization vector, and spatial direction of the thermally generated spin current, respectively. $l_{x(y)}$ is the length of the metallic film of the LSSE device along the $x$ ($y$) direction and $l_z$ is the thickness of the metallic film. (e) Schematic illustrations of the LSSE-device structures being developed for future thermoelectric applications. }\label{fig:introduction}
\end{center}
\end{figure*}

The Seebeck effect refers to the generation of an electric field ${\bf E}_{\rm SE}$ as a result of a temperature gradient $\nabla T$ in a conductor, where the direction of ${\bf E}_{\rm SE}$ is parallel to that of $\nabla T$ [Fig. \ref{fig:introduction}(a)]. The thermopower induced by the Seebeck effect is represented by the Seebeck coefficient: $\alpha \equiv {\bf E}_{\rm SE}/\nabla T$, which is equal to the ratio of the generated electric voltage to the applied temperature difference owing to the collinear orientation of ${\bf E}_{\rm SE}$ and $\nabla T$. A thermoelectric module based on the Seebeck effect usually consists of a number of $\Pi$-structured thermocouples, i.e., junctions of two materials with different Seebeck coefficients; the thermocouples are serially connected to enhance the thermoelectric output [Fig. \ref{fig:introduction}(b)]. The output voltage of the thermoelectric module can be enhanced in proportion to the number of the thermocouple elements, although such cascaded structure requires costly fabrication processes. \par
The efficiency of the thermoelectric generation based on the Seebeck effect is characterized by the dimensionless
figure of merit $Z_{\rm SE} T$, defined as
\begin{equation}\label{equ:ZT-Seebeck}
Z_{\rm SE} T = \frac{\alpha^2}{\kappa \rho}T,
\end{equation}
where $\kappa$, $\rho$, and $T$ are the thermal conductivity, electrical resistivity, and absolute temperature of thermoelectric materials \cite{DiSalvo1999Science}. The factor $\alpha^2/\rho$ is known as a thermoelectric power factor. The efficiency of the Seebeck device for electricity generation is defined as $\eta =$~(energy provided to a load)/(heat energy absorbed at the hot end of the device), which is a function of internal and load resistances. The optimized efficiency $\eta_{\rm SE}^*$ of the Seebeck device is then given as a function of the figure of merit as follows: 
\begin{equation}\label{equ:efficiency-Seebeck}
\eta_{\rm SE}^* = \frac{\Delta T}{T_{\rm h}} \frac{{ \sqrt{1 +Z_{\rm SE} \overline{T}} - 1 }}{{ \sqrt{1 +Z_{\rm SE} \overline{T}} + \frac{T_{\rm c}}{T_{\rm h}} }},
\end{equation}
where $\Delta T = T_{\rm h}-T_{\rm c}$ is the temperature difference between the hot and cold ends of the device, $T_{\rm h(c)}$ is the temperature at the hot (cold) end, and $\overline{T}$ is the average temperature between the hot and cold ends \cite{Rowe,Goldsmid}. Importantly, $\eta_{\rm SE}^*$ monotonically increases with increasing $Z_{\rm SE} \overline{T}$; $\eta_{\rm SE}^*$ goes to the Carnot efficiency $\eta_{\rm C} = \Delta T/T_{\rm h}$ in the limit of $Z_{\rm SE} \overline{T} \rightarrow \infty$. Therefore, many efforts in thermoelectric research are dedicated to improve the figure of merit of thermoelectric materials. \par
Equations (\ref{equ:ZT-Seebeck}) and (\ref{equ:efficiency-Seebeck}) mean that materials having a large Seebeck coefficient, low thermal conductivity, and low electrical resistivity are necessary for improving the thermoelectric conversion efficiency. Importantly, low $\kappa$ and $\rho$ values make it possible to suppress the energy loss due to heat conduction and Joule dissipation, respectively. However, in isotropic metals, the Wiedemann-Franz law ($\kappa_{\rm e}\rho = L_{\rm e} T$ with the electronic Lorenz number $L_{\rm e}$) limits this improvement when $\kappa$ is dominated by the electronic thermal conductivity $\kappa_{\rm e}$. A conventional way to overcome this limitation is to use thermoelectric semiconductors, where the thermal conductance is usually dominated by phonons while the electrical conductance is determined by charge carriers, and thus $\kappa$ and $\rho$ are separated according to the kind of the carriers \cite{DiSalvo1999Science}. Recently, to improve the figure of merit of thermoelectric materials by reducing phonon thermal conductivity without affecting electrical conductivity, not only exploration of new materials \cite{SnSe1,SnSe2} but also nanotechnology-based phonon engineering \cite{TE-nano1,TE-nano2,TE-nano3,TE-nano4,TE-nano5,TE-nano6,TE-nano7,TE-nano8,TE-nano9} have been conducted. Furthermore, research on the thermoelectric properties of anisotropic materials has attracted increasing attention recently \cite{SnSe1,SnSe2,Morelli_diacetylene,Balandin_graphene,Cohn2012PRL,Butler_2D-review,Fei_phosphorene}; since the Wiedemann-Franz law is violated in anisotropic materials, they are also useful for optimizing heat and charge conductions separately. \par
Nearly 200 years after the discovery of the Seebeck effect, a novel thermoelectric generation principle was discovered in the field of spintronics \cite{spintronics1,spintronics2,spincaloritronics-Bauer}. This novel principle can be said as a spin counterpart of the Seebeck effect: ``spin Seebeck effect'' (SSE) \cite{SSE_Uchida2008Nature}. The SSE refers to the generation of a spin current, a flow of spin angular momentum \cite{spincurrent1,spincurrent2}, as a result of a temperature gradient in a magnetic material. The SSE is applicable to the construction of thermoelectric generators because the spin current generated by the SSE can be converted into a charge current via the spin-orbit interaction, or the inverse spin Hall effect (ISHE) \cite{ISHE_Azevedo,ISHE_Saitoh,ISHE_Valenzuela,ISHE_Costache,ISHE_Kimura,ISHE_Sinova}, in a conductive thin film (mostly, a paramagnetic metal film) adjacent to the magnetic material (see Sec. \ref{sec:setup} for details). Here, the direction of the electric field induced by the ISHE ${\bf E}_{\rm ISHE}$ in the conductive film is perpendicular to that of $\nabla T$ in the magnetic material [Fig. \ref{fig:introduction}(c)], a configuration different from the conventional Seebeck effect. Since the SSE appears not only in ferromagnetic metals \cite{SSE_Uchida2008Nature,SSE_Uchida2010JAP,SSE_Bosu2011PRB,SSE_Uchida2011NatMat,TSSE_Wang2013PRB} and semiconductors \cite{SSE_Jaworski2010NatMat,SSE_Jaworski2011PRL} but also in magnetic (mostly, ferrimagnetic) insulators \cite{SSE_Uchida2010NatMat,SSE_Uchida2010APL_1}, it enables the conversion of heat energy in insulators into electrical energy in adjacent conductors \cite{SSE_Kirihara2012NatMat,SSE-Uchida2014JPCM}, which was impossible if only conventional thermoelectric technologies were used. The thermoelectric technology based on the SSE is still in an early phase of its development, and the efficiency is very small at the present stage. However, as reviewed in this article, the SSE exhibits various unconventional features suitable for thermoelectric applications. \par
\begin{figure*}[tb]
\begin{center}
\includegraphics{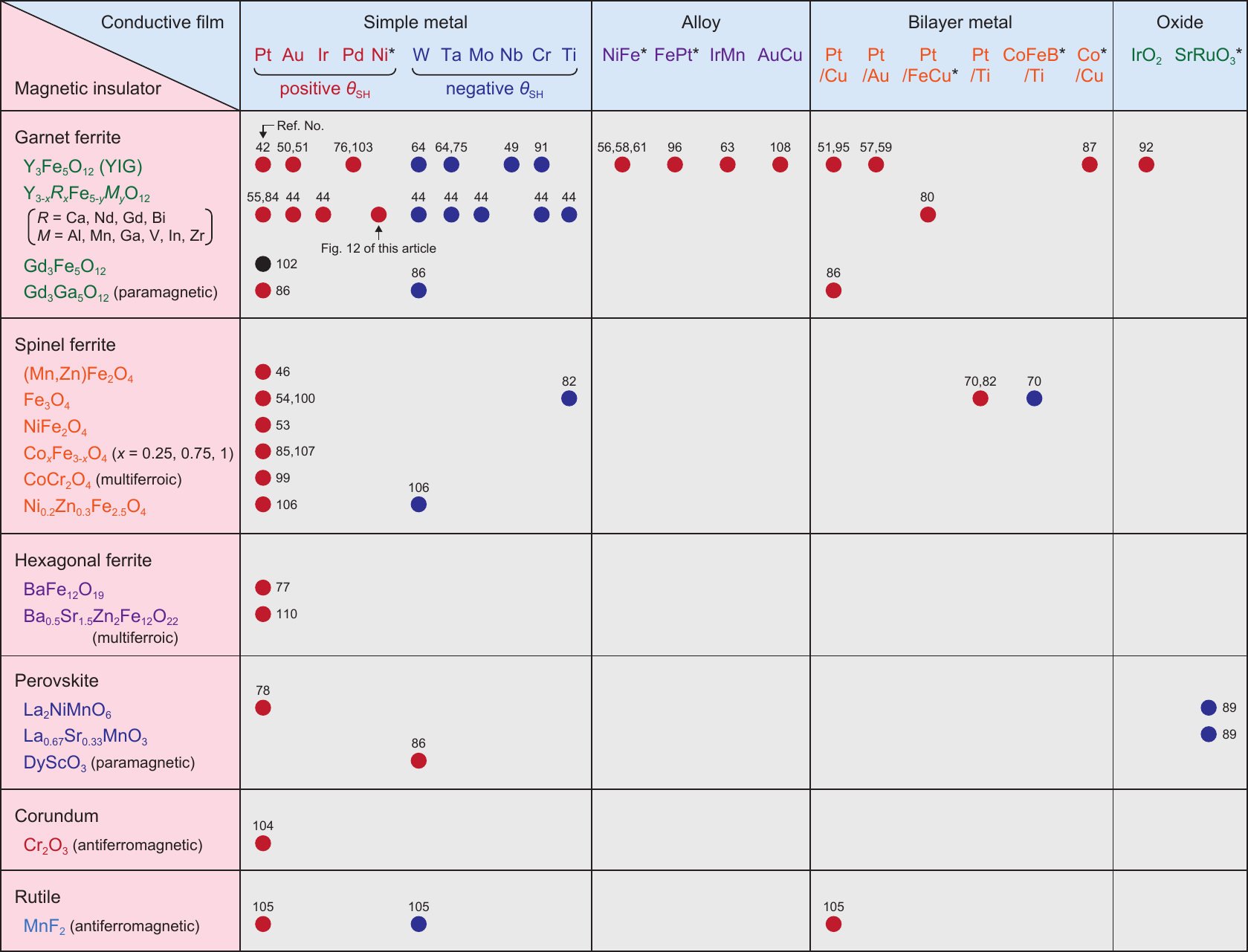}
\caption{Combination of magnetic insulators and conductive films used for the observation of the LSSE. The solid circles show the material combinations in which the LSSE have been observed, where the red (blue) circles represent the fact that the sign of the observed LSSE voltage is the same as (opposite to) that in the Pt/YIG systems. The color of the circle showing the Pt/Gd$_3$Fe$_5$O$_{12}$ system is black, since the LSSE in this system exhibits sign reversals as a function of temperature. The numbers near the circles correspond to the reference numbers. The asterisks in the list of the conductive films represent ferromagnetic metals, which enable the hybrid thermoelectric generation based on the LSSE and anomalous Nernst effect (ANE). Typical values for the spin diffusion lengths and spin Hall angles of the conductive materials are summarized in the review on the spin Hall effects \cite{ISHE_Sinova}. }\label{fig:LSSEmaterials}
\end{center}
\end{figure*}
In this review, we focus on the so-called longitudinal SSE (LSSE) \cite{SSE_Uchida2010APL_1,SSE-Uchida2014JPCM} in metallic film/magnetic insulator junction systems, depicted in Fig. \ref{fig:introduction}(c). After the first report on the SSE \cite{SSE_Uchida2008Nature}, this phenomenon has been measured mainly in two different device structures, i.e., the longitudinal and transverse configurations \cite{SSE_Uchida2012JAP}, where spin currents parallel and perpendicular to $\nabla T$ are measured, respectively. The longitudinal configuration has been used in most of the recent SSE studies because the insulator-based LSSE systems are useful for the exclusive detection of spin-current-induced signals and suitable for thermoelectric applications owing to its simple and versatile structure. The first observation of the LSSE was reported in 2010 by using a junction system comprising a paramagnetic metal Pt and a ferrimagnetic insulator Y$_3$Fe$_5$O$_{12}$ (YIG) \cite{SSE_Uchida2010APL_1}. The Pt/YIG junction system is now recognized as a model system for studying SSE physics, since Pt and YIG enable efficient spin-charge conversion and pure detection of spin-current effects, respectively. Since this demonstration, the LSSE has been observed in various combinations of magnetic insulators and conductive films \cite{SSE_Uchida2010APL_1,SSE_Kirihara2012NatMat,SSE-Uchida2014JPCM,SSE_Uchida2012JAP,SSE_MnZnFO,SSE-Uchida2011JJAP,SSE_Weiler2012PRL,SSE_Uchida2012APEX,SSE_Qu2013PRL,SSE_Kikkawa2013PRL,SSE_Jungfleisch2013APL,SSE_Meier2013PRB_NFO,SSE_Ramos2013APL_Fe3O4,SSE_Uchida2013PRB,SSE_Miao2013PRL_PyYIG,SSE_Schreier2013PRB,SSE_Kikkawa2013PRB,SSE_Schreier2013APL,SSE_Rezende2014PRB,SSE_Azevedo2014APL_PyYIG,SSE_BiYIG_Siegel,SSE_Mendes2014PRB_IrMn,SSE_Qu2014PRB,SSE Kehlberger2014JAP,SSE_time_resolved1,SSE_time_resolved2,SSE_time_resolved3,SSE_Saiga2014APEX,SSE_Wu2014APL_Fe3O4,SSE_Lustikova2014JAP,SSE_Aqeel2014JAP,SSE_Wang2014APL,SSE_Uchida2014PRX,SSE_Vlietstra2014PRB,SSE_Xu2014APL,SSE_BaFe12O19,SSE_La2NiMnO6,SSE_sign,SSE_BiYIG_Kikuchi,SSE_Qiu2015JPD,SSE_Wu2015JAP_Fe3O4,SSE_Sola2015JAP,SSE_Nd-YIG,SSE_Niizeki2015AIPAdv_CFO,SSE_Wu2015PRL,SSE_Tian2015APL,SSE_ML_Lee,SSE_LSMO_La2NiMnO6,SSE_Uchida2015PRB_PMA,SSE_Qu2015PRB_Cr,SSE_Qiu2015APEX,SSE_Kikkawa2015PRB,SSE_Jin2015PRB,SSE_Kehlberger2015PRL,SSE_TSeki2015APL,SSE_Wang2015Nanoscale,SSE_Kirihara2015IEEE,SSE_GdIG2016NatCommun,SSE_ML_Ramos,SSE_Ritzmann2015,SSE_Guo2015,SSE_CCO,SSE_Cr2O3,SSE_MnF2,SSE_Kirihara2015,SSE_Guo2016CFO,SSE_Zou2016PRB,SSE_Miao2016AIPAdv,SSE_Takagi2016BSZFO}, as shown in Fig. \ref{fig:LSSEmaterials}. In contrast, in the transverse configuration, the SSE measurements may be disturbed by thermal conductivity mismatch \cite{Huang2011PRL,Bosu2012JAP} between a film and a substrate that induces parasitic LSSE \cite{Meier2015NatCommun} or anomalous Nernst effect (ANE) \cite{spincaloritronics-Heremans,ANE_Lee2004PRL,ANE_Miyasato2007PRL,ANE_Pu2008PRL,ANE_Mizuguchi2012APEX,ANE_Sakuraba2013APEX,ANE_Ramos2014PRB,ANE_Sakuraba2016,ANE_Uchida2015}, requiring careful thermal design of the sample and measurement systems. In the transverse SSE measurements using conductive ferromagnets, a planar Nernst effect (PNE) \cite{spincaloritronics-Heremans,Avery2012PRL,Schmid2013PRL,Meier2013PRB,Soldatov2014PRB} may also arise. In the past several years, various thermoelectric measurements have been performed in the transverse configuration \cite{Huang2011PRL,Bosu2012JAP,Meier2015NatCommun,Avery2012PRL,Schmid2013PRL,Meier2013PRB,Soldatov2014PRB}, and the quantitative separation of the transverse SSE from the parasitic ANE and PNE has been reported in a ferromagnetic metal \cite{TSSE_Wang2013PRB}. In addition to the longitudinal and transverse configurations, a non-local geometry has also been used recently to investigate the length scale of the SSE \cite{Cornelissen2015NatPhys,Goennenwein2015APL,Giles2015PRB,Cornelissen2016PRB}. \par
The LSSE in insulators allows a new approach to improve the thermoelectric figure of merit. In the insulator-based LSSE device, the heat and charge currents flow in different parts of the device: $\kappa$ is the thermal conductivity of the magnetic insulator and $\rho$ is the electrical resistivity of the adjacent metal, such that $\kappa$ and $\rho$ in the LSSE device are segregated according to the part of the device elements \cite{SSE-Uchida2014JPCM}. Therefore, the denominator of the figure of merit, $\kappa \rho$, of the LSSE device is free from the Wiedemann-Franz law, and allowed to be optimized if one selects the combination of a magnetic insulator with low thermal conductivity and a metallic film with low electrical resistivity. This is one of the unique characteristics of the LSSE device, although, in addition to the $\kappa \rho$ factor, the LSSE thermopower itself must be improved. Importantly, in fact, the relation between the figure of merit and the thermoelectric conversion efficiency of the LSSE device should be different from that of the conventional Seebeck device due to the different device configurations and driving principles \cite{SSE_Cahaya2014APL,SSE_Liao2015Nanoscale,SSE_Cahaya2015IEEE}. One of the purposes of this review article is to formulate the thermoelectric figure of merit and conversion efficiency of the LSSE device, and to clarify the upper limit of the efficiency. \par
The advantages of the LSSE device include not only the device-design flexibility, discussed above, but also the following characteristics. The LSSE device has convenient scaling capability, where the thermoelectric output increases simply by extending the device area because the total amount of the thermally generated spin currents increases as the device becomes larger \cite{SSE_Kirihara2012NatMat}. Here, the electric  field induced by the ISHE at each point of the LSSE device is integrated into the output LSSE voltage: $V_{\rm SSE} = E_{\rm ISHE} l_y$, where $E_{\rm ISHE}$ is the magnitude of ${\bf E}_{\rm ISHE}$ and $l_y$ is the length of the metallic layer along the $y$ direction [Fig. \ref{fig:introduction}(c)]. Since the internal resistance of the metallic layer of the LSSE device is $R_{\rm metal} = \rho l_y/l_x l_z$ with $l_x$ and $l_z$ respectively being the length along the $x$ direction and thickness of the metallic film, the maximum extractable electrical power $P_{\rm max}$ is proportional to the area of the LSSE device: $P_{\rm max} \propto V_{\rm SSE}^2/R_{\rm metal} \propto l_x l_y$ [see Sec. \ref{sec:CVPcharacteristics_scaling}]. Such a straightforward scaling law makes it possible to enhance the thermoelectric output simply by enlarging the device area [Fig. \ref{fig:introduction}(d)]. To exploit the above characteristics of the LSSE device, versatile and low-cost fabrication methods have been developed, such as coating \cite{SSE_Kirihara2012NatMat} and plating \cite{SSE_Kirihara2015} technologies [Fig. \ref{fig:introduction}(e)]. These methods enable the implementation of simple-structure, large-area, and even flexible LSSE devices onto various heat sources, as shown in Secs. \ref{sec:coating} and \ref{sec:flexible}. \par
The thermoelectric generation based on the LSSE involves the following three factors: 
\begin{itemize}
 \item[(1)] heat-current/spin-current conversion efficiency in magnetic materials, 
 \item[(2)] spin-angular-momentum transfer efficiency across metal/insulator interfaces, characterized by the spin mixing conductance \cite{spin-mixing,Jia_spin-mixing_2011EPL,spin mixing concept,spin-mixing_Ohnuma}, 
 \item[(3)] spin-current/charge-current conversion efficiency in metallic films, characterized by the spin Hall angle \cite{ISHE_Sinova,Jiao2013PRL}. 
\end{itemize}
A direct approach to enhance the performance of the spin-current-driven thermoelectric generation is to improve the LSSE itself [the factor (1)], which realizes efficient thermal spin-current generation [see Sec. \ref{sec:multilayer}]. In addition to this approach, the improvement of the spin mixing conductance [the factor (2)] and the spin Hall angle [the factor (3)] are also essential. For example, to realize the efficient spin-angular-momentum transfer across the metal/insulator interface, the improvement of crystal quality and interface condition of the LSSE device by annealing has been conducted \cite{SSE_Saiga2014APEX,Benjamin_spin-pump_2013APL,Qiu_spin-pump_2013APL}. The spin mixing conductance can also be enhanced by inserting an ultra-thin ferromagnetic interlayer between a paramagnetic metal and a ferrimagnetic insulator owing to the increase of magnetic moment density at the interface \cite{SSE_BiYIG_Kikuchi,Jia_spin-mixing_2011EPL}. To realize the efficient spin-charge conversion, the ISHE has been measured in various metals \cite{SSE-Uchida2014JPCM,Morota_2011PRB,Wang2014PRL,Singh2015APL}, alloys \cite{SSE_Miao2013PRL_PyYIG,SSE_Kikkawa2013PRB,SSE_Azevedo2014APL_PyYIG,SSE_Mendes2014PRB_IrMn,SSE_Wu2014APL_Fe3O4,SSE_TSeki2015APL,SSE_Zou2016PRB,Niimi2011PRL,Niimi2012PRL,Laczkowski2014APL}, semiconductors \cite{Ando2011NatMat,Chen2013NatCommun,Sanchez2013PRB,Lee2014APL}, oxides \cite{SSE_LSMO_La2NiMnO6,SSE_Qiu2015APEX,Qiu2012APL,Fujiwara2013NatCommun}, and organic materials \cite{Ando2013NatMat,Qiu2015AIPAdv}. The spin mixing conductance and spin Hall angle can be estimated by means of various techniques, such as micro-wave-induced spin pumping \cite{ISHE_Azevedo,ISHE_Saitoh,Wang2014PRL,Singh2015APL,SSE_Azevedo2014APL_PyYIG,SSE_Mendes2014PRB_IrMn,Laczkowski2014APL,Ando2011NatMat,Chen2013NatCommun,Sanchez2013PRB,Lee2014APL,Qiu2012APL,Ando2013NatMat,Qiu2015AIPAdv,Silsbee1979PRB,Tserkovnyak2002PRL,Mizukami2002PRB,Kajiwara2010Nature}, spin Hall magnetoresistance \cite{Chen2013PRB,Nakayama2013PRL,Hahn2013PRB,Vliestra2013PRB,Althammer2013PRB}, and non-local methods \cite{ISHE_Valenzuela,ISHE_Kimura,Morota_2011PRB,Niimi2011PRL,Niimi2012PRL,Laczkowski2014APL,Fujiwara2013NatCommun,Jedema2001Nature,Jedema2002Nature,Niimi2015RepProgPhys}, where these phenomena can be measured by applying microwaves or charge currents to similar paramagnet/ferromagnet junction systems instead of temperature gradients. Therefore, not only the LSSE experiments but also these techniques are necessary for optimizing the LSSE devices. Although the conventional LSSE experiments have mainly been performed using paramagnetic metal/ferrimagnetic insulator junction systems, all-ferromagnetic devices \cite{SSE_Miao2013PRL_PyYIG,SSE_Kikkawa2013PRB,SSE_Azevedo2014APL_PyYIG,SSE_TSeki2015APL}, i.e., ferromagnetic metal/ferrimagnetic (or ferromagnetic) insulator junction systems, and alternately-stacked metal/insulator multilayer films \cite{SSE_ML_Ramos} have recently been recognized as useful tools for improving the thermoelectric performance of the LSSE devices, as shown in Secs. \ref{sec:ferro} and \ref{sec:multilayer}. \par
In this article, we review recent studies on the LSSE from the viewpoint of thermoelectric applications. This article is organized as follows. We start with an explanation of the thermoelectric generation process and measurement configuration of the LSSE in Sec. \ref{sec:setup}, followed by experimental demonstrations of the basic characteristics of the LSSE devices in Sec. \ref{sec:basic-characteristics}. In Sec. \ref{sec:calc-efficiency}, we formulate the thermoelectric figure of merit and conversion efficiency of the LSSE devices and compare the results with those for conventional thermoelectric devices. In Sec. \ref{sec:application}, we show preliminary demonstrations for future thermoelectric applications of the LSSE devices, where we focus on the device structures depicted in Fig. \ref{fig:introduction}(e). The last Sec. \ref{sec:conclusion} is devoted to the conclusions and prospects. \par
%
%
%%%%%%%%%%%%%%%%%%%%%%%%%%%%%%%%%%%%%%%%%%%%%%%%%%%%
\section{Thermoelectric generation process and measurement configuration} \label{sec:setup}
\begin{figure}[tb]
\begin{center}
\includegraphics{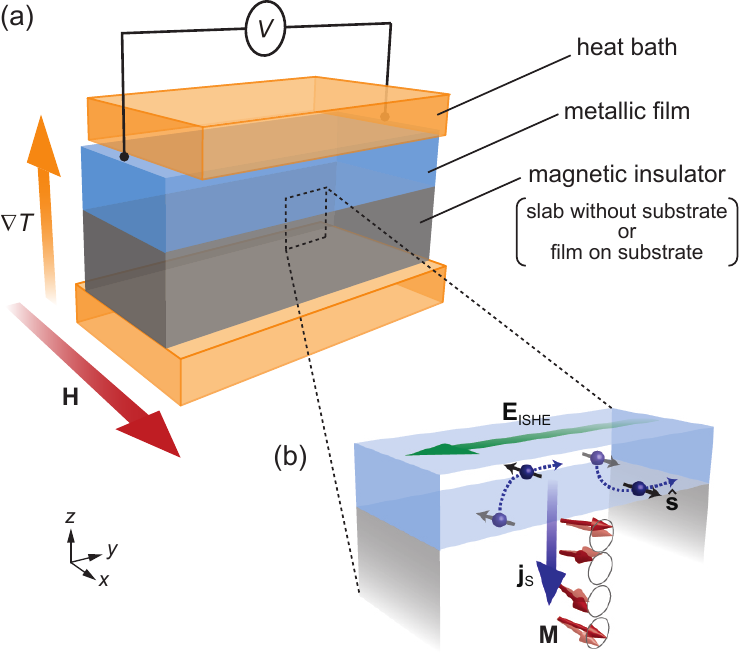}
\caption{(a) A schematic illustration of the typical experimental configuration for measuring the LSSE in a junction system comprising a metallic film and a magnetic insulator. $\nabla T$ is applied across the metal/insulator interface by sandwiching the LSSE device between two heat baths with different temperatures. When a soft magnetic insulator is used, an external magnetic field ${\bf H}$ (with the magnitude $H$) is applied to align the magnetization of the insulator along the $x$ direction. (b) A schematic illustration of the ISHE induced by the LSSE. $\hat{\bf s}$ denotes the unit vector along the electron-spin polarization ($||~{\bf M}$) in the metallic film. }\label{fig:LSSE-ISHE-setup}
\end{center}
\end{figure}

%%%%%%%%%%%%%%%%%%%%%%%%%%%%%%%%%%%%%%%%%%%%%%%%%%%%
%
%
Figure \ref{fig:LSSE-ISHE-setup}(a) shows a schematic illustration of the standard configuration used for the LSSE measurements. The basic structure of the LSSE device consists of a metallic film formed on a magnetic insulator. Here, the magnetic insulator is a slab without a substrate or a film formed on a substrate, while the metallic layer must be a thin film as explained later. In the metal/insulator junction in the longitudinal configuration, the temperature gradient, $\nabla T$, is applied perpendicular to the metal/insulator interface (along the $z$ direction) and the spatial direction of the spin current induced by the LSSE is parallel to the $\nabla T$ direction \cite{SSE_Uchida2010APL_1}. The LSSE is usually measured with applying an external magnetic field to align the magnetization of the magnetic insulator along the $x$ direction [Fig. \ref{fig:LSSE-ISHE-setup}(a)]. However, the external magnetic field is not indispensable for the LSSE-based thermoelectric generation if the device comprises a hard-magnetic material \cite{SSE_BaFe12O19,SSE_Niizeki2015AIPAdv_CFO}. Under this condition, a DC electric voltage difference $V$ between the ends of the metallic film of the LSSE device is measured. \par
To apply the temperature gradient in the direction perpendicular to the metal/insulator interface, the LSSE device is usually sandwiched between two heat baths which are stabilized at different temperatures [Fig. \ref{fig:LSSE-ISHE-setup}(a)] \cite{SSE_Uchida2010APL_1,SSE-Uchida2014JPCM}. By heating or cooling the heat baths, one can apply a reversible temperature gradient to the LSSE device since its top and bottom surfaces are directly connected to the heat bath \cite{SSE_Uchida2012JAP}. Owing to this high temperature controllability, the setup depicted in Fig. \ref{fig:LSSE-ISHE-setup}(a) is mostly used in the LSSE measurements; all the experiments shown in this article were performed in this setup. In contrast, the temperature gradient can be applied to the LSSE device also by using Joule heating in an on-chip heater \cite{SSE_Wu2014APL_Fe3O4,SSE_Vlietstra2014PRB,SSE_Xu2014APL,SSE_Wu2015JAP_Fe3O4}, laser heating \cite{SSE-Uchida2011JJAP,SSE_Weiler2012PRL,SSE_Schreier2013PRB,SSE_time_resolved1,SSE_time_resolved2}, and microwave heating \cite{SSE_Jungfleisch2013APL,SSE_time_resolved3}. Although these methods enable the measurements of the LSSE without using the specialized temperature-gradient generators, it is difficult to determine the temperature distributions; the methods are sometimes combined with numerical simulations \cite{SSE_Wu2015JAP_Fe3O4,SSE_Weiler2012PRL,SSE_Schreier2013PRB,SSE_time_resolved2,SSE_time_resolved3}. \par
The thermoelectric generation mechanism of the LSSE device is summarized as follows. The driving force of the LSSE is nonequilibrium dynamics of magnons, collective excitations of localized magnetic moments, in the magnetic insulator driven by the temperature gradient \cite{SSE_Xiao2010PRB,SSE_Adachi2010APL,SSE_Adachi2011PRB,SSE_Ohe2011PRB,SSE_Bender2012PRL,SSE_Ohnuma2013PRB,SSE_Adachi2013review,SSE_Hoffman2013PRB,SSE_Chotorlishvili2013PRB,SSE_Ren2013PRB,SSE_Bender2015PRB,SSE_Lyapilin2015PRB,SSE_Etesami2015APL,SSE_Brataas2015PRB,SSE_Chotorlishvili2015JMMM,SSE_Rezende2016PRB,SSE_Xiao2015}, since the LSSE appears even when a conduction electrons' contribution is completely frozen out. This nonequilibrium magnon dynamics in the magnetic insulator can interact with conduction-electron spins in the attached metal and transfer the spin angular momentum across at the metal/insulator interface via the interface s-d exchange interaction, which is described in terms of the spin mixing conductance \cite{spin-mixing,Jia_spin-mixing_2011EPL,spin mixing concept,spin-mixing_Ohnuma}. This spin-angular-momentum transfer induces a conduction-electron spin current in the metallic film. This spin current is converted into the electric field, ${\bf E}_{\rm ISHE}$, by the aforementioned ISHE in the metallic layer if the spin-orbit interaction of the metal is strong [Fig. \ref{fig:LSSE-ISHE-setup}(b)]. When the magnetization ${\bf M}$ of the magnetic insulator is aligned along the $x$ direction, ${\bf E}_{\rm ISHE}$ is generated in the metallic film along the $y$ direction according to the relation \cite{spincurrent1}
\begin{equation}\label{equ:ISHE}
{\bf E}_{\rm ISHE} = (\theta_{\rm SH} \rho ) \hat{\bf s} \times {\bf j}_{\rm S},
\end{equation}
where $\theta_{\rm SH}$, $\hat{\bf s}$, and ${\bf j}_{\rm S}$ are the spin Hall angle, unit vector along the electron-spin polarization ($||~{\bf M}$) in the metallic layer, and spatial direction of the spin current (with the magnitude of spin current density $j_{\rm S}$) induced by the LSSE, respectively. Thus, the thermoelectric voltage can be generated by the combination of the LSSE and ISHE and observed by measuring the voltage $V$ between the ends of the metallic layer, where the sign of $V$ is determined by that of $\theta_{\rm SH}$ (Fig. \ref{fig:LSSEmaterials}) \cite{SSE-Uchida2014JPCM}. This thermoelectric generation mechanism works below the Curie temperature of the magnetic insulator \cite{SSE_Uchida2014PRX}. Here, to detect the LSSE voltage, the thickness of the metallic layer has to be comparable to its spin diffusion length \cite{ISHE_Sinova,Jiao2013PRL}, typically a few to several hundreds of nanometers, since the spin current injected into the metallic layer and the resultant ${\bf E}_{\rm ISHE}$ exist only in the vicinity of the metal/insulator interface due to the spin relaxation. We also note that, when a highly resistive insulator is used, the contribution from thermoelectric phenomena in itinerant magnets, such as the conventional Seebeck and Nernst effects, is eliminated in the LSSE device \cite{SSE_Uchida2010APL_1,SSE_Kikkawa2013PRB}. \par
%
%
%%%%%%%%%%%%%%%%%%%%%%%%%%%%%%%%%%%%%%%%%%%%%%%%%%%%
\section{Basic characteristics of spin Seebeck devices} \label{sec:basic-characteristics}
%%%%%%%%%%%%%%%%%%%%%%%%%%%%%%%%%%%%%%%%%%%%%%%%%%%%
%
%
In this section, by using the Pt/YIG model systems, we demonstrate the basic characteristics of the LSSE devices, such as the dependence of the thermoelectric voltage on the temperature, temperature gradient and its direction, magnetic field and its direction, and device geometry. \par
\subsection{Fundamental experiments} \label{sec:fundamental-exp}
\begin{figure*}[bt]
\begin{center}
\includegraphics{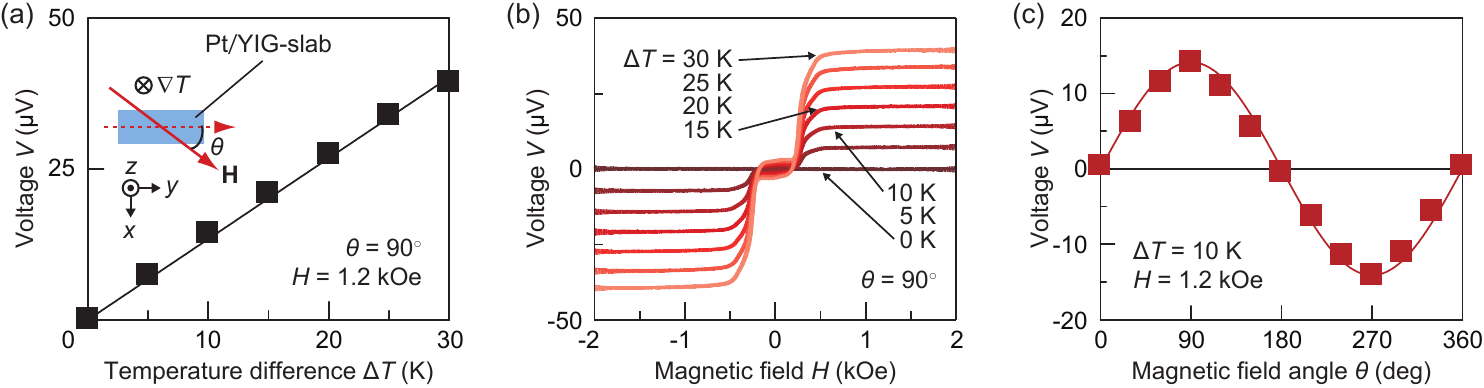}
\caption{(a) The temperature-difference $\Delta T$ dependence of the voltage $V$ in the Pt/YIG-slab sample at $H = 1.2~\textrm{kOe}$ and $\theta = 90^\circ$, measured when $\nabla T~|| -z$. $\theta$ denotes the angle between ${\bf H}$ and the $y$ direction. (b) $H$ dependence of $V$ in the Pt/YIG-slab sample for various values of $\Delta T$ at $\theta = 90^\circ$. (c) $\theta$ dependence of $V$ in the Pt/YIG-slab sample at $\Delta T = 10~\textrm{K}$ and $H = 1.2~\textrm{kOe}$. }\label{fig:fundamental_data1}
\end{center}
\end{figure*}
Here, we show the fundamental properties of the LSSE \cite{SSE-Uchida2014JPCM,SSE_Kikkawa2013PRL,SSE_Kikkawa2013PRB}. The Pt/YIG junction system used here consists of a single-crystalline YIG slab and a Pt film sputtered on the well-polished surface of the YIG. The lengths of the Pt film (YIG slab) along the $x$, $y$, and $z$ directions are $l_x = 2~\textrm{mm}$ ($L_x = 2~\textrm{mm}$), $l_y = 6~\textrm{mm}$ ($L_y = 6~\textrm{mm}$), and $l_z = 10~\textrm{nm}$ ($L_z = 1~\textrm{mm}$), respectively. To generate $\nabla T$ along the $z$ direction, the temperatures of the heat baths attached to the top and bottom of the Pt/YIG-slab sample were stabilized to $300~\textrm{K}$ and $300~\textrm{K}+\Delta T$, respectively. The external magnetic field ${\bf H}$ (with the magnitude $H$) was applied to the Pt/YIG-slab sample in the $x$-$y$ plane at an angle $\theta$ to the $y$ direction [see the inset to Fig. \ref{fig:fundamental_data1}(a)]. \par
Figure \ref{fig:fundamental_data1}(a) shows $V$ between the ends of the Pt film in the Pt/YIG-slab sample as a function of $\Delta T$ at $H=1.2\,\textrm{kOe}$. When ${\bf H}$ was applied along the $x$ direction ($\theta = 90^\circ$), the magnitude of $V$ was found to be proportional to $\Delta T$. As shown in Fig. \ref{fig:fundamental_data1}(b), the sign of $V$ for finite values of $\Delta T$ is reversed in response to the sign reversal of ${\bf H}$, indicating that the $V$ signal in the Pt film is affected by the magnetization direction of the YIG slab (note that the magnetization of the YIG slab is aligned along the ${\bf H}$ direction when $H > 0.5\,\textrm{kOe}$). To confirm the origin of this signal, we also measured the $\theta$ dependence of $V$ in the same Pt/YIG-slab sample at $\Delta T = 10~\textrm{K}$ and $H=1.2~\textrm{kOe}$ [Fig. \ref{fig:fundamental_data1}(c)]. The $V$ signal was observed to vary with $\theta$ in a sinusoidal pattern and vanish when $\theta = 0$ and $180^\circ$, a situation consistent with the symmetry of the ISHE induced by the LSSE described in Eq. (\ref{equ:ISHE}). \par
\begin{figure*}[tb]
\begin{center}
\includegraphics{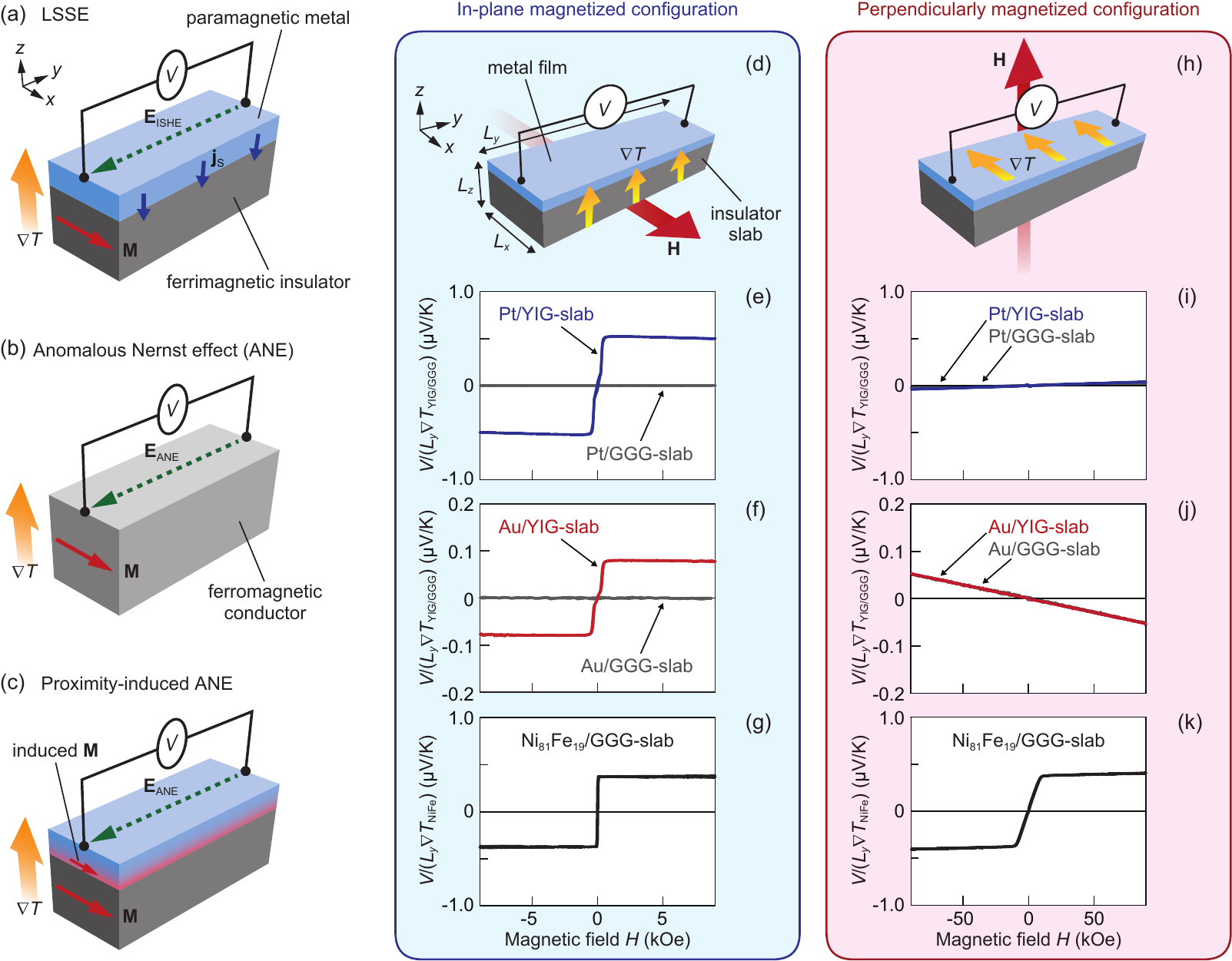}
\caption{(a)-(c) Schematic illustrations of the LSSE (a), ANE (b), and proximity-induced ANE (c). ${\bf E}_{\rm ANE}$ denotes the electric field generated by the ANE. (d) A schematic illustration of the metal-film/insulator-slab junction system in the in-plane magnetized (IM) configuration. (e),(f) $H$ dependence of $V/(L_y \nabla T_{\rm YIG(GGG)})$ in the Pt/YIG (GGG)-slab and Au/YIG (GGG)-slab samples in the IM configuration. (g) $H$ dependence of $V/(L_y \nabla T_{\rm NiFe})$ in the Ni$_{81}$Fe$_{19}$/GGG-slab sample in the IM configuration. (h) A schematic illustration of the metal-film/insulator-slab junction system in the perpendicularly magnetized (PM) configuration. (i),(j) $H$ dependence of $V/(L_y \nabla T_{\rm YIG(GGG)})$ in the Pt/YIG (GGG)-slab and Au/YIG (GGG)-slab samples in the PM configuration. (k) $H$ dependence of $V/(L_y \nabla T_{\rm NiFe})$ in the Ni$_{81}$Fe$_{19}$/GGG-slab sample in the PM configuration. $L_{x}$, $L_y$, and $L_{z}$ are the lengths of the insulator slab along the $x$, $y$, and $z$ directions, respectively. $\nabla T_{\rm YIG(GGG)}$ and $\nabla T_{\rm NiFe}$ denote the temperature gradients in the YIG (GGG) slab and Ni$_{81}$Fe$_{19}$ film, respectively. The sample consists of a 10-nm-thick Pt, Au, or Ni$_{81}$Fe$_{19}$ film formed on a single-crystalline YIG or GGG slab. All the measurements were performed at $T = 300~\textrm{K}$. }\label{fig:fundamental_data2}
\end{center}
\end{figure*}
The above experiments indicate that the observed thermoelectric voltage is attributed to the LSSE. However, to exclusively establish the LSSE, the contribution of the LSSE has to be separated from that of the ANE \cite{spincaloritronics-Heremans}. Since Pt is a paramagnetic metal and YIG is a very good insulator, the ANE does not exist in the Pt/YIG system in the ordinary sense. In this system, however, weak ferromagnetism may be induced in the Pt layer in the vicinity of the Pt/YIG interface due to a static magnetic proximity effect \cite{Huang2012PRL,Geprags_2012APL_XMCD,Chien_2013PRL_XMCD,Geprags_2013arXiv_XMCD,Sun2013PRL,Miao2014PRL,proximity-ANE_Guo,Haertinger2015PRB} because Pt is near the Stoner ferromagnetic instability \cite{DoS,Ibach}. In fact, when the thickness of Pt is very thin ($< 3~\textrm{nm}$), weak ferromagnetic signals were observed by means of X-ray magnetic circular dichroism \cite{Geprags_2012APL_XMCD,Chien_2013PRL_XMCD,Geprags_2013arXiv_XMCD}. If the proximity-induced ferromagnetism induces the ANE in the Pt layer, the ISHE voltage induced by the LSSE in the Pt/YIG system may be contaminated by the proximity-induced ANE in the Pt layer. This possibility was pointed out by Huang {\it et al.} in 2012 on the basis of magnetoresistance and Hall measurements \cite{Huang2012PRL}, although the anisotropic magnetoresistance in Pt/YIG systems was subsequently shown to be attributed to the spin Hall magnetoresistance \cite{Chen2013PRB,Nakayama2013PRL,Hahn2013PRB,Vliestra2013PRB,Althammer2013PRB}. Here, the electric field induced by the ANE is generated according to the relation \cite{SSE-Uchida2014JPCM}
\begin{equation}\label{equ:ANE1}
{\bf E}_{\rm ANE} = {\cal N} {\bf M} \times \nabla T, 
\end{equation}
where ${\cal N}$ is the anomalous Nernst coefficient. This configuration is similar to that of the LSSE since ${\bf E}_{\rm ANE}$ is generated along the $y$ direction when $\nabla T~||~z$ and ${\bf M}~||~x$ [compare Figs. \ref{fig:fundamental_data2}(a)-\ref{fig:fundamental_data2}(c)]. After the problem presentation by Huang {\it et al.}, the pure detection of the LSSE was realized by using Au/YIG systems \cite{SSE_Qu2013PRL,SSE_Kikkawa2013PRL}, where Au is free from the magnetic proximity effect because its electronic structure is far from the Stoner instability \cite{DoS}. To investigate the magnetic proximity effect in Pt/YIG systems, microwave spectroscopy measurements were also carried out \cite{Sun2013PRL,Haertinger2015PRB}. Guo {\it et al.} theoretically investigated the proximity-induced ANE in Pt and Pd within Berry-phase formalism based on relativistic band-structure calculations \cite{proximity-ANE_Guo}. \par
The clear separation of the LSSE from the proximity-induced ANE was reported in \cite{SSE_Kikkawa2013PRL,SSE_Kikkawa2013PRB} by comparing transverse thermoelectric voltage in the Pt/YIG system in in-plane magnetized (IM) and perpendicularly magnetized (PM) configurations. In the IM (PM) configuration, ${\bf H}$ is applied parallel (perpendicular) to the Pt/YIG interface and $\nabla T$ is applied perpendicular (parallel) to the interface, as shown in Fig. \ref{fig:fundamental_data2}(d) [\ref{fig:fundamental_data2}(h)]. The IM configuration is the same as the LSSE setup (Fig. \ref{fig:LSSE-ISHE-setup}), where both the LSSE and ANE can appear if they exist. In the PM configuration, the ANE signal can appear since the temperature gradient, magnetization, and inter-electrode direction are at right angles to one another [Eq. (\ref{equ:ANE1})], while the LSSE voltage should disappear due to the symmetry of the ISHE [Eq. (\ref{equ:ISHE})], where $\hat{\bf s}~||~{\bf j}_{\rm S}$ in the PM configuration. Therefore, the quantitative comparison of the voltage between these configurations enables the estimation of the ANE contamination in the Pt/YIG system. \par
Figures \ref{fig:fundamental_data2}(e) and \ref{fig:fundamental_data2}(i) show the $H$ dependence of the voltage normalized by the device length along the $y$ direction and the temperature gradient in the YIG slab, $V/(L_y \nabla T_{\rm YIG})$, in the Pt/YIG-slab sample in the IM and PM configurations, respectively \cite{SSE-Uchida2014JPCM,SSE_Kikkawa2013PRB}. The magnitude of $V/(L_y \nabla T_{\rm YIG})$ in the IM configuration was found to be much greater than that in the PM configuration. Here, the magnitude of the normal Nernst voltage, which is the $H$-linear component of $V/(L_y \nabla T_{\rm YIG})$, in the Pt/YIG-slab sample in the PM configuration is comparable to that in the Pt/paramagnetic Gd$_3$Ga$_5$O$_{12}$ (GGG)-slab sample and in a Pt plate without a substrate \cite{SSE_Kikkawa2013PRB}, confirming that the in-plane temperature gradient is generated in the Pt/YIG-slab sample in the PM configuration (note that the Pt/GGG-slab sample exhibits only a normal Nernst effect and no LSSE voltage except at very low temperatures \cite{SSE_Wu2015JAP_Fe3O4}). The voltage behavior in the Pt/YIG-slab sample is completely different from that in a ferromagnetic Ni$_{81}$Fe$_{19}$ film on a GGG substrate, where only the ANE and small normal Nernst effect appear; in the Ni$_{81}$Fe$_{19}$ film, the isotropic ANE voltage was observed in both the IM and PM configurations, when the voltage is normalized by the temperature gradient in the Ni$_{81}$Fe$_{19}$ film [Figs. \ref{fig:fundamental_data2}(g) and \ref{fig:fundamental_data2}(k)] \cite{SSE-Uchida2014JPCM,SSE_Kikkawa2013PRB}. The above results clearly show that the transverse thermoelectric voltage in the Pt/YIG system is dominated by the ISHE voltage induced by the LSSE and that the proximity-ANE contamination is negligibly small. In \cite{SSE_Kikkawa2013PRB}, the contribution of the proximity-induced ANE voltage in the Pt/YIG system was estimated to be less than 0.1~\% of the LSSE voltage. As shown in Figs. \ref{fig:fundamental_data2}(f) and \ref{fig:fundamental_data2}(j), similar results were obtained in Au/YIG systems. \par
The experimental results shown in Fig. \ref{fig:fundamental_data2} clearly demonstrate that the thermoelectric voltage in the Pt/YIG system is due entirely to the ISHE induced by the LSSE. Although the separation of the LSSE from the ANE is very important from the viewpoint of fundamental physics, even the proximity-induced ANE should be utilized effectively from the viewpoint of thermoelectric applications because the thermoelectric output of the LSSE device might be enhanced by the superposition of the proximity-induced ANE voltage. Recently, to validate the availability of the proximity-induced ANE, we investigated the ANE in alternately-stacked Pt/Fe multilayer films, which were designed to enhance the proximity-induced ANE intentionally; the proximity-ANE contribution in the Pt/Fe multilayer films is expected to be up to two orders of magnitude greater than that in the conventional Pt/YIG system \cite{ANE_Uchida2015}. However, even in these multilayer systems, no clear evidence for the existence of the proximity-induced ANE was observed. Therefore, the potential of the magnetic proximity effect as a thermoelectric generation mechanism is still unknown. In contrast, the standard ANE in ferromagnetic materials has already been investigated as a novel thermoelectric conversion technology \cite{ANE_Sakuraba2013APEX,ANE_Sakuraba2016}. Especially, we focus on ferromagnetic metals as conductive layer materials of the LSSE device since they enable the hybrid thermoelectric generation by the LSSE and ANE \cite{SSE-Uchida2014JPCM} and replacement of noble metals, such as Pt and Au, in the LSSE device, as discussed in Sec. \ref{sec:ferro}. \par
\subsection{High magnetic field dependence}
\begin{figure*}[tb]
\begin{center}
\includegraphics{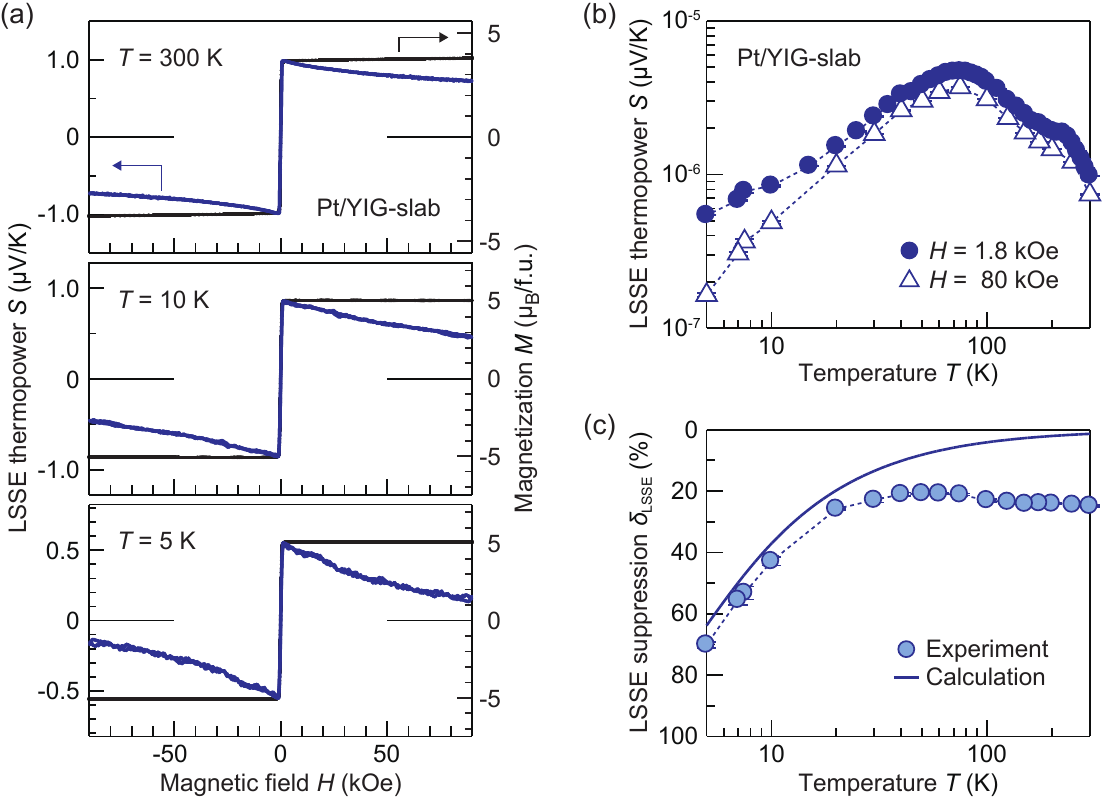}
\caption{(a) $H$ dependence of the LSSE thermopower $S$ in the Pt/YIG-slab sample and the magnetization $M$ of the YIG slab at $T = 300~\textrm{K}$, 10 K, and 5 K, measured when $H$ was swept between $\pm 90~\textrm{kOe}$. The sample consists of a 5-nm-thick Pt film sputtered on the top surface of a single-crystalline YIG slab. (b) Double logarithmic plot of the $T$ dependence of $S$ in the Pt/YIG-slab sample at $H =$ 1.8 kOe (closed circles) and 80 kOe (open triangles). (c) $T$ dependence of the suppression of the LSSE voltage by magnetic fields $\delta _{\rm SSE}$ in the Pt/YIG-slab sample (circles). A blue line shows the $T$ dependence of $\delta _{\rm SSE}$ calculated based on the conventional SSE model (see \cite{SSE_Kikkawa2015PRB} for details). }\label{fig:high-field1}
\end{center}
\end{figure*}
Recently, several research groups reported the investigation of the high-magnetic-field response of the LSSE in Pt/YIG systems with different YIG thicknesses at various temperatures \cite{SSE_Kikkawa2015PRB,SSE_Jin2015PRB,SSE_Ritzmann2015,SSE_Guo2015}. These studies provide a significant guideline for optimizing the thickness of the magnetic insulator layer of the LSSE device since the high-magnetic-field dependence of the LSSE was found to be associated with the characteristic lengths of the LSSE. In this subsection, with showing our experimental results for the Pt/YIG systems \cite{SSE_Kikkawa2015PRB}, we review the behavior of the LSSE in a high magnetic field range and its interpretation. \par
The observation of the SSE in insulators revealed that the magnon excitation plays a key role in this phenomenon. After the pioneering theoretical work by Xiao {\it et al.} \cite{SSE_Xiao2010PRB}, the SSE is mainly described in terms of the effective magnon temperature $T_{\rm m}$ in a ferrimagnetic insulator and effective electron temperature $T_{\rm e}$ in an attached paramagnetic metal; when the effective magnon-electron temperature difference is induced by an external temperature gradient, a spin current is generated across the ferrimagnet/paramagnet interface due to the thermal spin pumping. Subsequently, Adachi {\it et al.} developed linear-response theories of the magnon- and phonon-mediated SSEs \cite{SSE_Adachi2010APL,SSE_Adachi2011PRB,SSE_Adachi2013review}. Hoffman {\it et al.} formulated a Landau-Lifshitz-Gilbert theory of the SSE to investigate the thickness dependence and length scale of the SSE \cite{SSE_Hoffman2013PRB}. In 2014, Rezende {\it et al.} discussed the SSE in terms of a bulk magnon spin current created by a temperature gradient in a ferrimagnetic insulator \cite{SSE_Rezende2014PRB}. Furthermore, various theoretical models of the magnon-driven SSE were also developed \cite{SSE_Ohe2011PRB,SSE_Bender2012PRL,SSE_Ohnuma2013PRB,SSE_Chotorlishvili2013PRB,SSE_Ren2013PRB,SSE_Bender2015PRB,SSE_Lyapilin2015PRB,SSE_Etesami2015APL,SSE_Brataas2015PRB,SSE_Chotorlishvili2015JMMM,SSE_Rezende2016PRB,SSE_Xiao2015}. However, microscopic understanding of the relation between the magnon excitation and thermally generated spin current is yet to be fully established, and more detailed studies are necessary. Since the magnon excitation is modulated by a magnetic field due to the Zeeman gap $g\mu_{\rm B}H$, the ISHE voltage induced by the SSE can also be affected by the magnetic field. Therefore, systematic measurements of the magnetic-field-induced response of the SSE become powerful tools for unraveling the thermo-spin conversion mechanism based on the magnon excitation. \par
In Fig. \ref{fig:high-field1}(a), we show the transverse thermopower $S~(\equiv (V/\Delta T)(L_z / L_{y}))$ in the Pt/YIG-slab sample as a function of $H$ for several values of the temperature $T$, measured when $H$ was swept between $\pm 90~\textrm{kOe}$. The clear LSSE voltage was observed in the Pt/YIG-slab sample at all the temperatures and its magnitude at each temperature gradually decreases with increasing $H$ after taking the maximum value, while the magnitude of $M$ is almost constant after the saturation. This suppression of the LSSE voltage becomes apparent by applying high magnetic fields, while it is very small in the conventional LSSE measurements in a low field range. The suppression of the LSSE voltage was shown to be irrelevant to the normal Nernst effect in the Pt film \cite{SSE_Kikkawa2015PRB}. \par
The magnetic-field-induced suppression of the LSSE voltage in the Pt/YIG-slab sample increases with decreasing the temperature. Figure \ref{fig:high-field1}(c) shows the $T$ dependence of the suppression of the LSSE thermopower $\delta _{\rm SSE}$ in the same Pt/YIG-slab sample, where $\delta _{\rm SSE}$ is defined as $(S_{\rm max} - S_{80\textrm{kOe}})/S_{\rm max}$ with $S_{\rm max}$ and $S_{80\textrm{kOe}}$ respectively being the $S$ values at the maximum point and at $H = 80~\textrm{kOe}$ [the $T$ dependence of $S$ at the positive $H$ values is shown in Fig. \ref{fig:high-field1}(b)]. The field-induced suppression in the Pt/YIG-slab sample was observed to be almost constant above 30 K and strongly enhanced below 30 K; the $\delta _{\rm SSE}$ value in the Pt/YIG-slab sample reaches $\sim$70 \% at $T = 5~\textrm{K}$. The field-induced suppression of the LSSE for $T > 30~\textrm{K}$ cannot be explained by the conventional SSE models, while that for $T < 30~\textrm{K}$ seemingly agrees with numerical calculations based on the thermal spin pumping mechanism [Fig. \ref{fig:high-field1}(c)] (see \cite{SSE_Kikkawa2015PRB} for details). The inconsistency between the observed suppression of the LSSE voltage and the conventional formulation at relatively-high temperatures is attributed to the fact that the small Zeeman energy is defeated by thermal fluctuations when $g\mu_{\rm B}H \ll k_{\rm B} T$ in the conventional models, where $g$, $\mu_{\rm B}$, and $k_{\rm B}$ are the $g$-factor, Bohr magneton, and Boltzmann constant, respectively (note that the magnon gap energy at $H = 80~\textrm{kOe}$ corresponds to $g\mu_{\rm B}H/k_{\rm B}= 10.7~\textrm{K}$); to affect the magnon excitation by magnetic fields, the magnon energy has to be comparable to or less than the Zeeman energy. In contrast, the observed large suppression of the LSSE voltage in the Pt/YIG-slab sample indicates that the magnon excitation relevant to the LSSE is affected by magnetic fields even at around room temperature. This result suggests that low-frequency magnons of which the energy is comparable to the Zeeman energy provide a dominant contribution to the LSSE; the thermo-spin conversion efficiency of the LSSE has magnon-frequency dependence, which is not included in the conventional SSE theories. In \cite{SSE_Kikkawa2015PRB} and \cite{SSE_Rezende2016JMMM}, the origin of this spectral non-uniform thermo-spin conversion is discussed in terms of the frequency dependence of a magnon thermalization (energy relaxation) length and a magnon diffusion length, respectively. It is notable that lower frequency magnons exhibit the longer characteristic lengths in general \cite{SSE_Xiao2010PRB,SSE_Rezende2016JMMM,Sanders-Walton,SSE_Zhang-Zhang_2012PRL,SSE_Zhang-Zhang_2012PRB,Agrawal2013PRL,Ruckriegel2014PRB,YIG_kappa_Boona-Heremans,Cornelissen2015NatPhys,Goennenwein2015APL,Giles2015PRB,Cornelissen2016PRB}. \par
\begin{figure}[tb]
\begin{center}
\includegraphics{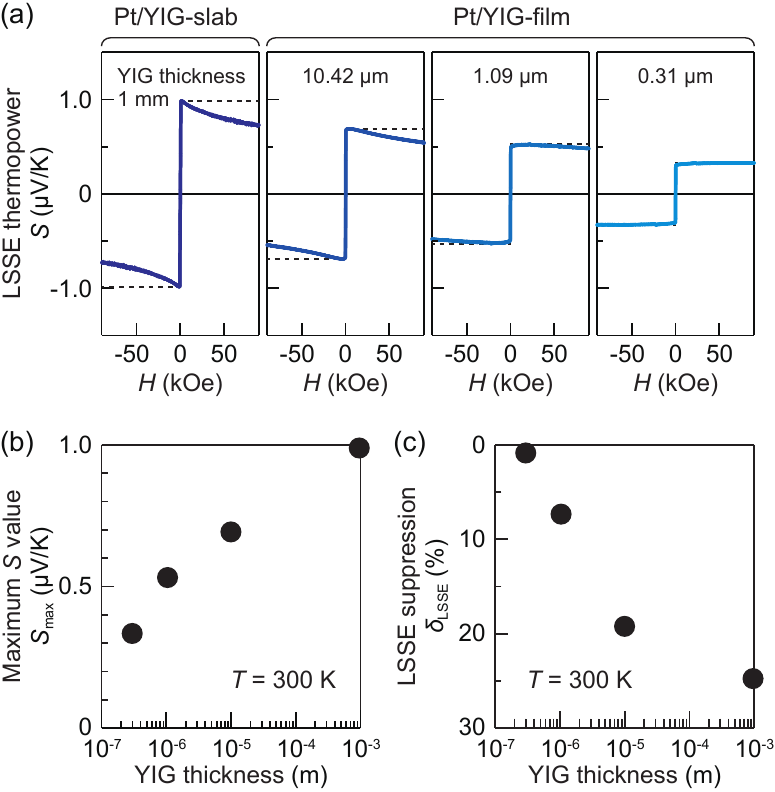}
\caption{(a) $H$ dependence of $S$ in the Pt/YIG-slab sample with the YIG thickness of $1~\textrm{mm}$ and in the Pt/YIG-film samples with the YIG thickness of $10.42~\mu \textrm{m}$, $1.09~\mu \textrm{m}$, and $0.31~\mu \textrm{m}$ at $T = 300~\textrm{K}$, measured when $H$ was swept between $\pm 90~\textrm{kOe}$. The YIG films were grown on single-crystalline GGG substrates by a liquid phase epitaxy method. (b) YIG-thickness dependence of $S_{\rm max}$ at $T = 300~\textrm{K}$. (c) YIG-thickness dependence of $\delta _{\rm SSE}$ at $T = 300~\textrm{K}$. }\label{fig:high-field2}
\end{center}
\end{figure}
The above results and discussions indicate that, to maximize the thermoelectric output of the LSSE device, the thickness of the magnetic insulator has to be greater than the characteristic lengths of low-frequency magnons providing a strong contribution to the LSSE, since the contribution from the long-range magnons can be limited by boundary conditions in thin magnetic insulators \cite{SSE_Kikkawa2015PRB}. In fact, several research groups demonstrated that, by using the Pt/YIG-slab and Pt/YIG-film systems, the magnitude of the LSSE thermopower monotonically decreases with decreasing the thickness of YIG [our experimental results are shown in Figs. \ref{fig:high-field2}(a) and \ref{fig:high-field2}(b)] \cite{SSE_Kikkawa2015PRB,SSE_Kehlberger2015PRL,SSE_Guo2015}. Significantly, the suppression of the LSSE by high magnetic fields, $\delta _{\rm SSE}$, also monotonically decreases with decreasing the YIG thickness [Figs. \ref{fig:high-field2}(a) and \ref{fig:high-field2}(c)]. This behavior indicates that the contribution of low-frequency magnons, which govern the LSSE suppression in the Pt/YIG-slab sample, fades away in the Pt/YIG-film samples when the YIG thickness is less than their characteristic lengths and that only remaining contribution from high-frequency magnons, which have energy much greater than the Zeeman energy and provide a weak contribution to the LSSE, appears in the thin YIG-film samples. As reviewed in this subsection, the measurements of the high-magnetic-field response of the LSSE are useful for characterizing the properties of the LSSE devices associated with magnon excitation. \par
\subsection{Current-voltage-power characteristics and scaling law} \label{sec:CVPcharacteristics_scaling}
\begin{figure*}[tb]
\begin{center}
\includegraphics{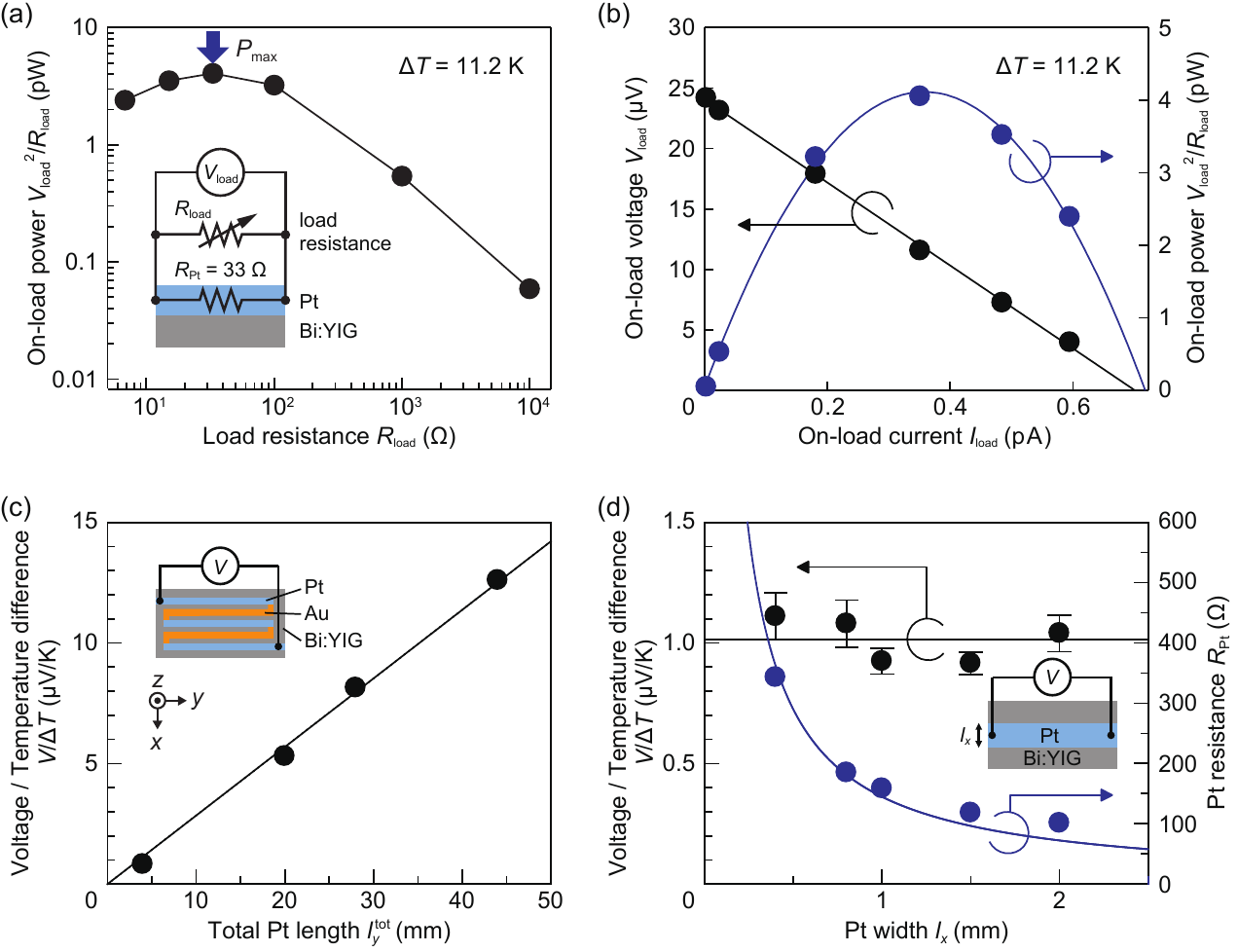}
\caption{(a) On-load power $V_{\rm load}^2/R_{\rm load}$ induced by the LSSE in the Pt/Bi:YIG-film sample as a function of the load resistance $R_{\rm load}$ at $\Delta T = 11.2~\textrm{K}$. $V_{\rm load}$ denotes the on-load voltage. (b) $V_{\rm load}$ and $V_{\rm load}^2/R_{\rm load}$ induced by the LSSE in the Pt/Bi:YIG-film sample as a function of the on-load current $I_{\rm load} = V_{\rm load}/R_{\rm load}$. The Pt/Bi:YIG-film sample used for the experiments (a) and (b) consists of a 10-nm-thick Pt film with the size of $16 \times 16~\textrm{mm}^2$ and a 120-nm-thick Bi:YIG film formed on a 0.5-mm-thick SGGG wafer of 1-inch diameter. (c) $V/\Delta T$ as a function of the total Pt length $l_{y}^{\rm tot}$ in the [Pt-Au thermopile]/Bi:YIG-film samples. (d) $V/\Delta T$ and the Pt-film resistance $R_{\rm Pt}$ as a function of the Pt width $l_{x}$ in the Pt/Bi:YIG-film samples. All the measurements were performed at room temperature. }\label{fig:CVPcharacteristics_scaling}
\end{center}
\end{figure*}
Now we demonstrate the current-voltage-power characteristics of the LSSE device by using a Pt/BiY$_2$Fe$_5$O$_{12}$ (Bi:YIG) bilayer film \cite{SSE-Uchida2014JPCM}. To determine the characteristics, we attached a load resistance $R_{\rm load}$ to the Pt layer and measured the voltage across the load resistance $V_{\rm load}$ to estimate the output power $V_{\rm load}^2/R_{\rm load}$ with changing the $R_{\rm load}$ value, while most of the LSSE experiments have been performed under the open-circuit condition. In Fig. \ref{fig:CVPcharacteristics_scaling}(a), we show $V_{\rm load}^2/R_{\rm load}$ generated from the Pt/Bi:YIG-film sample at $\Delta T = 11.2~\textrm{K}$ as a function of $R_{\rm load}$. We found that the maximum output power can be extracted when $R_{\rm load}$ is equal to the internal resistance of the Pt layer: $R_{\rm Pt} = 33~\Omega$. This behavior is also confirmed by the linear current-voltage relation and parabolic current-power relation for the Pt/Bi:YIG-film sample in Fig. \ref{fig:CVPcharacteristics_scaling}(b). The current-voltage-power characteristics and optimizing condition for the load resistance for the LSSE devices are the same as those for any linear generators, such as batteries and conventional thermoelectric modules. \par
Next, we experimentally verify the scaling law of the LSSE device: the output power is proportional to the device area, introduced in Sec. \ref{sec:intro}. To do this, we investigated the dependence of the LSSE voltage on the size, i.e., length and width, of the metallic layer of the LSSE device as follows \cite{SSE_Kirihara2015IEEE}. \par
For measuring the metallic-layer-length dependence of the LSSE, we used the structure called ``spin Hall thermopile'' \cite{SSE_Uchida2012APEX}. The spin Hall thermopile consists of an alternating array of two different metallic wires with different spin Hall angles, connected in series in a zigzag configuration. The sample used here is the spin Hall thermopile comprising 10-nm-thick Pt wires and 30-nm-thick Au wires formed on a Bi:YIG film [see the inset to Fig. \ref{fig:CVPcharacteristics_scaling}(c)]. In this sample, the LSSE voltage in each wire is integrated into the total voltage between the ends of the Pt-Au thermopile and the total voltage predominantly comes from the Pt wires, because the spin Hall angle and resistivity of Pt are greater than those of Au and the thickness of the Pt wires is less than that of the Au wires (note that the magnitude of the LSSE voltage increases with decreasing the thickness of the metallic layer except for ultrathin regions \cite{SSE_Kikkawa2013PRB,SSE_Qu2014PRB,SSE_Saiga2014APEX}). Figure \ref{fig:CVPcharacteristics_scaling}(c) shows the dependence of the LSSE voltage per unit temperature difference, $V/\Delta T$, on the total length of the Pt wires $l_{y}^{\rm tot}$ in the Pt-Au thermopile, where $l_{y}^{\rm tot}$ is changed by changing the number of the Pt-Au pairs. As expected above, the magnitude of $V/\Delta T$ increases in proportion to $l_{y}^{\rm tot}$. \par
In addition to the Pt-length dependence, we also investigated the Pt-width dependence. We fabricated Pt films with different widths on fixed-size Bi:YIG films, where the length and thickness of the Pt films are fixed [see the inset to Fig. \ref{fig:CVPcharacteristics_scaling}(d)]. In Fig. \ref{fig:CVPcharacteristics_scaling}(d), $V/\Delta T$ and the Pt-film resistance $R_{\rm Pt}$ are plotted as a function of the Pt width $l_x$. We found that the magnitude of $V/\Delta T$ is almost independent of $l_x$, while $R_{\rm Pt}$ is inversely proportional to $l_x$. \par
The obtained scaling results, i.e., $V/\Delta T \propto l_{y}^{\rm tot}$ and $V/(R_{\rm Pt} \Delta T) \propto l_x$, clearly indicate that a larger film area leads to larger thermoelectric output power that can be extracted to an external load. Here, we would like to emphasize again that the scaling law of the LSSE devices is quite different from that of conventional thermoelectric devices based on the Seebeck effect, where the thermoelectric output scales with the number of thermocouples connected in series (see Sec. \ref{sec:intro} and Fig. \ref{fig:introduction}). \par
%
%
%%%%%%%%%%%%%%%%%%%%%%%%%%%%%%%%%%%%%%%%%%%%%%%%%%%%
\section{Theory of efficiency of spin Seebeck thermoelectric devices} \label{sec:calc-efficiency}
%%%%%%%%%%%%%%%%%%%%%%%%%%%%%%%%%%%%%%%%%%%%%%%%%%%%
%
%
In this section, we present a theory of the efficiency of thermoelectric devices based on the LSSE. First, we give a brief review of the efficiency of thermoelectric generators based on the ANE. This is because the ANE has a similarity to the LSSE in that the induced electric field is orthogonal to the applied temperature gradient, and a thermoelectric device possessing this characteristic is called a transverse device~\cite{Goldsmid}. Next, based on the discussion of the ANE generators, we formulate the efficiency of the LSSE devices. \par
\subsection{Review of efficiency of thermoelectric generators based on anomalous Nernst effects} 
As pointed out above, it is quite instructive to review the efficiency calculation of the ANE device~\cite{Harman62,Harman62-2,Harman63}. We begin with the following transport equations~\cite{Landau-Elec}: 
%%%   
\begin{eqnarray}
  {\bf E} &=& \rho {\bf j} + \alpha {\bf \nabla}T + {\cal N} {\bf M} \times {\bf \nabla}T, 
  \label{Eq:ANEtrans01a}\\
  {\bf q} &=& \Pi {\bf j} -\kappa {\bf \nabla}T + {\cal N} T {\bf M} \times {\bf j}, 
  \label{Eq:ANEtrans01b} 
\end{eqnarray}
%%%
where ${\bf E}$, ${\bf j}$, and ${\bf q}$ are an applied electric field, charge current density (with the magnitude $j$), and heat current density, respectively. $\Pi= \alpha T$ is the Peltier coefficient. The (anomalous) Hall coefficient is disregarded as it is irrelevant to the present discussion, and the Leduc-Righi coefficient \cite{thermal-Hall1,thermal-Hall2} is assumed to be negligibly small. Here we note that the following discussion is applicable not only to the ANE devices but also to normal Nernst devices \cite{spincaloritronics-Heremans} if the spontaneous magnetization is replaced with an external magnetic field, while this article focuses on the comparison between the LSSE and ANE devices because of their similar characteristics. It should also be added that the thermoelectric conversion efficiency of normal Nernst generators, or Ettingshausen coolers, in semimetals can be much greater than that of the ANE and LSSE devices, although an external magnetic field must be applied \cite{Yim72}. \par
We first calculate the amount of heat evolved per unit time and volumes, which governs the temperature profile of the system in the steady state. This quantity is given by $- {\bf \nabla} \cdot {\bf q}_E$~\cite{Landau-Elec}, where ${\bf q}_E= {\bf q}+ \phi {\bf j}$ is the energy current density with $\phi$ being the electrochemical potential. Using Eqs.~(\ref{Eq:ANEtrans01a}) and (\ref{Eq:ANEtrans01b}), we obtain 
%%% 
\begin{eqnarray}
  -{\bf \nabla} \cdot {\bf q}_E = 
  \rho j^2 &+& \kappa \nabla^2 T \nonumber \\
  &-& T {\bf j} \cdot {\bf \nabla} \alpha +2{\cal N} ({\bf j} \times {\bf M}) \cdot {\bf \nabla} T, 
  \label{Eq:divQE01} 
\end{eqnarray}
%%% 
where the first and second terms on the right hand side represent the Joule heating and thermal conduction. The third term, which comes from the temperature dependence of the Seebeck coefficient via the relation ${\bf \nabla} \alpha = (d \alpha/dT) {\bf \nabla} T$, is the Thompson effect and will be neglected hereafter to simplify the argument. Besides, since the last term can be interpreted as a change in the Thompson effect due to the presence of magnetization~\cite{Landau-Elec}, this term is also discarded. Therefore, by considering the conservation law of energy flux, i.e., $\nabla \cdot {\bf q}_E =0$, we obtain the following equation:
%%%
\begin{equation}
  \rho j^2 + \kappa \nabla^2 T = 0, 
  \label{Eq:Domenicali01}
\end{equation}
%%%
which determines the temperature distribution in thermoelectric devices, known as Domenicali's equation~\cite{Domenicali54}. \par
\begin{figure*}[tb]
\begin{center}
\includegraphics{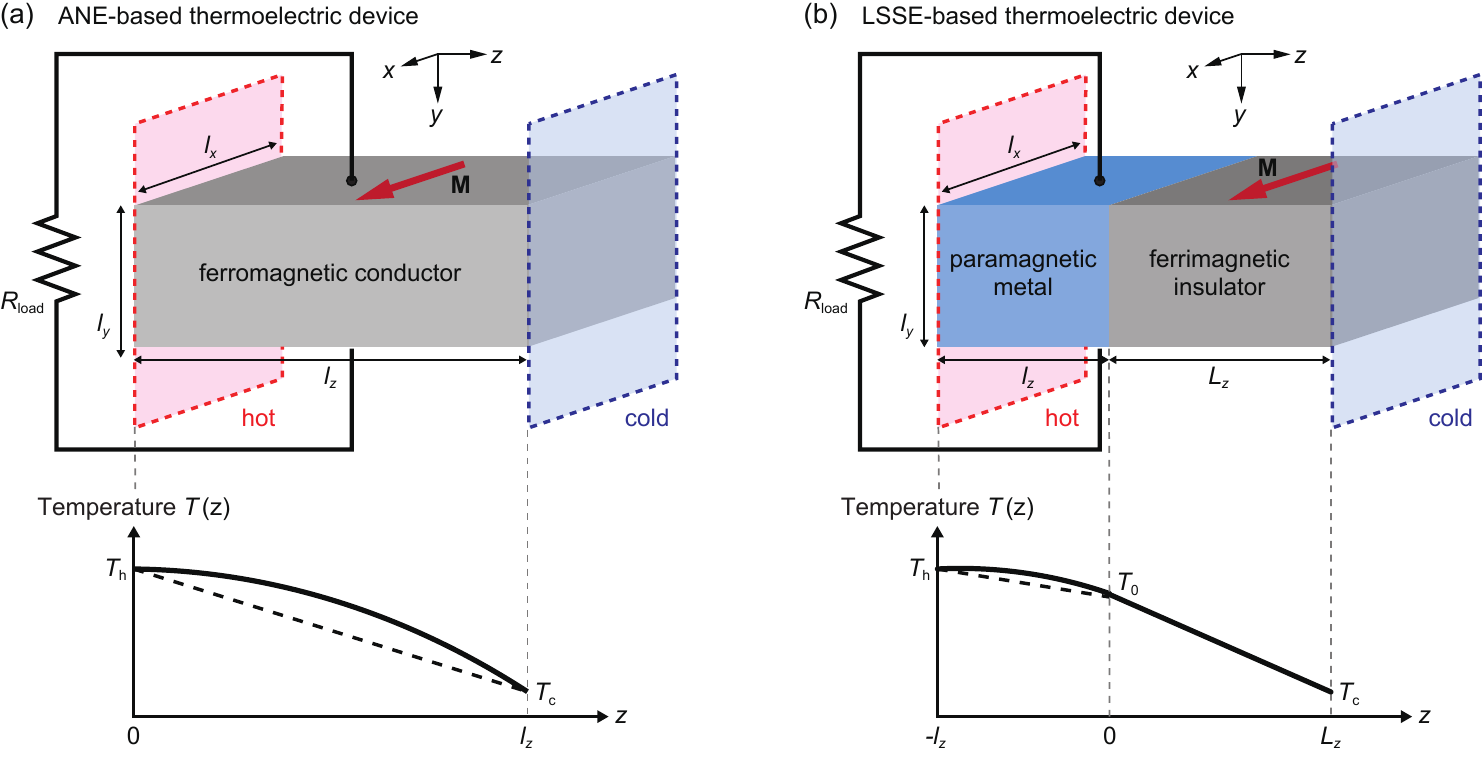}
\caption{(a) A schematic illustration of the ANE-based thermoelectric device and its temperature profile along the $z$ direction. (b) A schematic illustration of the LSSE-based thermoelectric device and its temperature profile along the $z$ direction. The load resistance $R_{\rm load}$ is attached to the ferromagnetic conductor (paramagnetic metal) of the ANE (LSSE) device. $l_x$, $l_y$, and $l_z$ are the lengths of the ferromagnetic conductor (paramagnetic metal) of the ANE (LSSE) device along the $x$, $y$, and $z$ directions, respectively. $T_{\rm h(c)}$ is the temperature at the hot (cold) end of the devices and $T_0$ is the temperature at the metal/insulator interface of the LSSE device. }\label{fig:calc-efficiency}
\end{center}
\end{figure*}
Let us focus on the ANE device shown in Fig.~\ref{fig:calc-efficiency}(a). Here, a temperature gradient $\nabla_z T= -\Delta T/l_z$ ($\Delta T= T_{\rm h}- T_{\rm c}$) is applied along the $z$ axis, and an external magnetic field is applied along the $x$ axis to align the magnetization, ${\bf M} = M_{\rm s} {\bf \hat{\bf x}}$. Under the isothermal condition ($\nabla_y T=0$) as well as the open circuit condition ($j= 0$), the ANE induces an electric field $E_y= {\cal N} M_{\rm s} \Delta T/l_z$ along the $y$ axis, and a charge current density is driven in the $y$ direction, i.e., ${\bf j} =j_y {\bf \hat{\bf y}}$. Since the electromotive force is given by 
%%%
\begin{equation}
V_{\rm emf} = -{\cal N} M_{\rm s} \Delta T (l_y/l_z), 
\end{equation}
%%%
and the total resistance in the circuit is given by $R_{\rm tot}= R+ R_{\rm load}$ with $R$ and $R_{\rm load}$ respectively being the internal and load resistances, the resultant charge current ${\bf I}= I_y {\bf \hat{\bf y}}$ is calculated to be 
%%%
\begin{eqnarray}
  I_y &=& \frac{V_{\rm emf}}{R_{\rm tot}} = \frac{-{\cal N} M_{\rm s} \Delta T(l_y/l_z)}{R(1+r)}, 
  \label{Eq:I_ANE01} 
\end{eqnarray}
%%%
where we have introduced a dimensionless variable $r \equiv R_{\rm load}/R$. \par
Now we discuss the efficiency of the ANE device. The efficiency is defined by 
%%%
\begin{equation}
  \eta_{\rm ANE} = P_{\rm out}/{Q}_{\rm h},  
  \label{Eq:eta_ANE01}
\end{equation}
%%%
where $P_{\rm out}= R_{\rm load} I_y^2$ is the electrical power output and ${Q}_{\rm h}$ is the thermal power input from the hot reservoir. The latter quantity is calculated as follows~\cite{Heikes61,Mahan98}. First, we assume a uniform charge current distribution along the $z$ direction, i.e., $j_y(z) = j= \text{constant}$. Then, from Domenicali's equation [Eq.~(\ref{Eq:Domenicali01})] under the condition $T(z=0)=T_{\rm h}$ and $T(z=l_z)=T_{\rm c}$, the temperature distribution is calculated to be 
%%%
\begin{eqnarray}
T(z) &=& T_{\rm h}-(z/l_z)\Delta T + (\rho j^2/2 \kappa) z(l_z -z), 
\end{eqnarray}
%%%
where the distribution of $T(z)$ is schematically shown in Fig.~\ref{fig:calc-efficiency}(a). Next, we evaluate the magnitude of the heat current at the hot reservoir, i.e., $q_{\rm h}=  {\cal N} M_{\rm s} T j - \kappa (\nabla_z T)_{z=0}$, using Eq.~(\ref{Eq:ANEtrans01b}). Then, the rate of heat removal from the hot reservoir, defined by ${Q}_{\rm h}=(l_x l_y) q_{\rm h}$, is calculated to be 
%%%
\begin{eqnarray}
  {Q}_{\rm h} &=& \frac{{\cal N} M_{\rm s} T_{\rm h}}{(l_z/l_y)} I_y 
  + K \Delta T - \frac{R}{2}I_y^2,   \label{Eq:Qh_ANE01}
\end{eqnarray}
%%%
where $K= \kappa (l_x l_y/l_z)$ is the thermal conductance in the $z$ direction and $R= \rho (l_y/l_x l_z)$ is the electrical resistance in the $y$ direction. \par
From Eqs.~(\ref{Eq:eta_ANE01})-(\ref{Eq:Qh_ANE01}), the efficiency is expressed as a function of $r=R_{\rm load}/R$: 
%%%
\begin{equation}
\eta_{\rm ANE}= \eta_{\rm C} f(r), 
\end{equation}
%%%
where $\eta_{\rm C}= \Delta T/T_{\rm h}$ is the Carnot efficiency. Here, the characteristic function, 
%%%
\begin{equation}
f(r)=\frac{r}{-(1+r) + \frac{(1+r)^2}{Z'_{\rm ANE} T_{\rm h}} - \frac{\Delta T}{2 T_{\rm h}} },  
\label{Eq:func_character}
\end{equation}
%%%
is defined by the isothermal figure of merit 
%%%
\begin{equation}
  Z'_{\rm ANE} = \frac{{({\cal N} M_{\rm s})^2}}{{\kappa' \rho}}, 
  \label{Eq:ZANE}
\end{equation} 
%%%
where $\kappa'$ is the thermal conductivity at zero electric field. Note the minus sign in the first term of the denominator in Eq.~(\ref{Eq:func_character}), which is peculiar to transverse devices~\cite{Harman62}. By maximizing $\eta_{\rm ANE}$ with respect to $r$, the optimized efficiency $\eta_{\rm ANE}^*$ in the limit of $\Delta T \ll T_{\rm h}$ is calculated to be 
%%%
\begin{eqnarray}
  \eta_{\rm ANE}^* &=& \eta_{\rm C} \left( \frac{1- r^*}{1+ r^* } \right), \\
  \label{Eq:etamax_ANE01}
  r^* &=& \sqrt{1- Z'_{\rm ANE} T_{\rm h} }, 
  \label{Eq:smax01}
\end{eqnarray}
%%%
where the functional form is quite different from the conventional Seebeck device because of the fact that the ANE device has a symmetry of transverse devices. Note that, by contrast, the maximum power output is given by the impedance matching condition $R=R_{\rm load}$ as is well known \cite{Heikes61}. \par
Two comments are in order. First, according to the argument of \cite{Horst63} and \cite{Delves64}, the isothermal figure of merit $Z'_{\rm ANE} T$ in Eq.~(\ref{Eq:func_character}) is related to the adiabatic figure of merit $Z_{\rm ANE} T= ({\cal N} M_{\rm s})^2/\kappa \rho$ as 
%%%
\begin{equation}
  Z'_{\rm ANE} T 
  = 
  \frac{Z_{\rm ANE} T}{1+Z_{\rm ANE} T} ,  
\label{eq:ZTrange}
\end{equation}
%%%
such that $0 \le Z'_{\rm ANE} T \le 1$, where $\kappa$ is the thermal conductivity at zero charge current. Therefore, the maximum allowed efficiency is obtained for $Z'_{\rm ANE} T=1$ (i.e., $r^*=0$), achieving the Carnot efficiency: $\eta_{\rm ANE}^{\rm max} = \eta_{\rm C}$. Second, the figure of merit $Z'_{\rm ANE}$ in Eq.~(\ref{Eq:ZANE}) is determined by the thermal conductivity $\kappa'$ and the electrical resistivity $\rho$ that are defined in the same material. \par
%
%Therefore, its maximization is limited by the Wiedemann-Franz law ($\kappa'_{\rm e} \rho=\text{constant}$), where $\kappa'_{\rm e}$ is the electronic part of thermal conductivity. 
%
\subsection{Efficiency of thermoelectric generators based on spin Seebeck effects} 
In this subsection, based on the discussion in the previous subsection, we discuss the efficiency of the LSSE device. We consider a device shown in Fig.~\ref{fig:calc-efficiency}(b), where a bilayer of a paramagnetic metal P and a ferrimagnetic insulator F is sandwiched between hot and cold reservoirs. We assume that the thermal conductivity of P $\kappa_{\rm P}$ is much greater than that of F $\kappa_{\rm F}$, such that the temperature gradient develops inside F and thus the temperature gradient inside P is negligibly small in comparison to F. Looking at Table I given in \cite{SPE_Flipse2014PRL}, this assumption seems moderately reasonable. Note that we made this assumption in order to keep the mathematical manipulations manageable and to extract a simple analytical expression; otherwise such a simple result would never be obtained and a more involved numerical approach would be required, which is beyond our scope. Also, we assume that there is no discontinuous jump in the temperature distribution at the P/F interface. In this sense, the present approach is complementary to the calculation in \cite{SSE_Cahaya2014APL}, where only the temperature difference at the P/F interface was considered as a driving force. Instead, we assume that the temperature is a smooth function across the P/F interface, and the temperature at the interface $T_0$ is determined by the heat current conservation. Here, the relationship between the present continuous temperature model and a three temperature model used in \cite{SSE_Schreier2013PRB,SSE_Cahaya2014APL,SSE_Xiao2010PRB} is given in the following way. The present approach is constructed based on the formalism developed in \cite{SSE_Zhang-Zhang_2012PRB}, and the magnon accumulation [see Eq. (11) of \cite{SSE_Zhang-Zhang_2012PRB}] corresponds to the effective magnon-phonon temperature difference. It has been demonstrated that the present single temperature gradient model provides us with an excellent description of the SSE in magnetic multilayers \cite{SSE_ML_Ramos}. \par
We start our discussion from the following transport equations for P: 
%%%
\begin{eqnarray}
  {\bf j} &=& \sigma_{\rm P} {\bf E} + \theta_{\rm SH} {\bf \hat{\bf s}} \times {\bf j}_{\rm S}, 
  \label{Eq:jc_N01} \\
  {\bf j}_{\rm S} &=& \sigma_{\rm P} {\bf E}_{\rm S}
  + \theta_{\rm SH} {\bf \hat{\bf s}} \times {\bf j}, 
  \label{Eq:js_N01}\\
  {\bf q} &=& - \kappa_{\rm P} {\bf \nabla}T, 
  \label{Eq:q_N01}
\end{eqnarray}
%%%
where ${\bf E}_{\rm S}$ is the spin voltage gradient and $\sigma_{\rm P}$ is the electrical conductivity of P. For F, we assume the following transport equations: 
%%%
\begin{eqnarray}
  {\bf j}_{\rm S} &=& \sigma_{\rm m} {\bf E}_{\rm S}-  \zeta_{\rm m} {\bf \nabla} T, 
  \label{Eq:js_mag01}\\
  {\bf q} &=& \Pi_{\rm m} {\bf j}_{\rm S} - \kappa_{\rm F} {\bf \nabla} T, 
  \label{Eq:q_mag01}
\end{eqnarray}
%%%
where ${\bf E}_{\rm S}$ in F is determined by the magnon density gradient ${\bf \nabla} n_{\rm m}$, $\sigma_{\rm m}$ is the magnon conductivity, and $\zeta_{\rm m}$ ($\Pi_{\rm m}$) is the magnon Seebeck (Peltier) coefficient. Here, the thermal conductivity $\kappa_{\rm F} = \kappa_{\rm F}^{\rm mag}+\kappa_{\rm F}^{\rm ph}$ of F includes both magnon contribution $\kappa_{\rm F}^{\rm mag}$ as well as phonon contribution $\kappa_{\rm F}^{\rm ph}$, and Onsager's reciprocity \cite{SPE_Flipse2014PRL} requires $\Pi_{\rm m} = \zeta_{\rm m} T/\sigma_{\rm m}$. \par
Let us first calculate the open circuit voltage under a temperature bias $\Delta T = T_{\rm h}- T_{\rm c}$ between the hot and cold reservoirs. As stated before, because of the condition $\kappa_{\rm P} \gg \kappa_{\rm F}$, the temperature gradient develops inside F as ${\bf \nabla}T = -[(T_{\rm h}- T_0)/L_z] {\bf \hat{\bf z}} \approx -(\Delta T/L_z) {\bf \hat{\bf z}}$. According to Eq.~(\ref{Eq:js_mag01}), this temperature gradient drives a spin current ${\bf j}_{\rm S} = \zeta_{\rm m} (\Delta T/L_z ) {\bf \hat{z}}$ inside F, which then injects an amount of 
%%%
\begin{eqnarray}
  {\bf j}_{\rm S} &=& c_1 \zeta_{\rm m} (\Delta T/L_z ) {\bf \hat{z}} 
  \label{Eq:js_SSE01} 
\end{eqnarray}
%%%
into P, where the coefficient $c_1$ ($0<c_1<1$) represents the ratio of the spin current injected into P to the thermal drift magnon current in F. Note that the parameter $c_1$ is proportional to the spin mixing conductance. If we adopt an approach of \cite{SSE_Rezende2014PRB} and \cite{SSE_Zhang-Zhang_2012PRB} where the continuity of ${\bf j}_{\rm S}$ at the P/F interface is postulated, the constant $c_1$ is given by $c_1 \propto [\cosh(L_z/\Lambda)-1]/\sinh(L_z/\Lambda)$, where $\Lambda$ is the magnon diffusion length of F. Inside P, thanks to the ISHE [Eq.~(\ref{Eq:jc_N01})], the injected spin current ${\bf j}_{\rm S}= [c_1 (\zeta_{\rm m} \Delta T)/L_z ] {\bf \hat{\bf z}}$ is converted into a charge current ${\bf j} = - [\theta_{\rm SH} c_1 \zeta_{\rm m} \Delta T/L_z] {\bf \hat{\bf y}}$ when ${\bf \hat{\bf s}} = {\bf \hat{\bf x}}$. Then, in the open circuit condition, this current is compensated by an electromotive force 
%%%
\begin{eqnarray}
  V_{\rm emf} &=& - \theta_{\rm SH} \alpha_{\rm S} \Delta T (L_y/L_z) , 
  \label{Eq:SSE_v0} 
\end{eqnarray}
%%%
where $\alpha_{\rm S}= (1/2) c_1 \zeta_{\rm m}/\sigma_{\rm P}$. Here, the factor $1/2$ in $\alpha_{\rm S}$ comes from averaging over the thickness of P, where we approximated the hyperbolic variation of $j_{\rm S} (z)$ by a linear curve by assuming that the thickness of P is of the order of its spin diffusion length $\lambda$ ($\sim l_z$). For a circuit containing the total resistance $R_{\rm tot}= R+ R_{\rm load}$, the current ${\bf I}= I_y {\bf \hat{\bf y}}$ driven by the open circuit voltage [Eq.~(\ref{Eq:SSE_v0})] is given by 
%%%
\begin{eqnarray}
  I_y &=& - \frac{\theta_{\rm SH} \alpha_{\rm S} \Delta T (L_y/L_z)}{R(1+r)}
\end{eqnarray}
%%%
where $r= R_{\rm load}/R$ as before. Owing to Eq.~(\ref{Eq:js_N01}), the resultant charge current density ${\bf j} = [I_y/(l_x l_z)] {\bf \hat{y}}$ drives an additional spin current $\delta {\bf j}_{\rm S} =  [(\theta_{\rm SH} I_y)/(l_x l_z)] {\bf \hat{z}} $, a part of which is injected back into F. Then, due to Eq.~(\ref{Eq:q_mag01}), this spin current is accompanied by a heat current 
%%%
\begin{eqnarray}
  \delta {\bf q} &=& 
  c_2 \frac{ \Pi_{\rm m} \theta_{\rm SH} I_y}{(l_x l_z)} {\bf \hat{z}}, 
\end{eqnarray}
%%%
where the coefficient $c_2$ ($0<c_2<1$) represents the ratio of the spin current injected back into F to the spin-Hall drift current $\delta j_{\rm S}$ in P. Note that the heat current accompanied by the spin current given in Eq.~(\ref{Eq:js_SSE01}), i.e., ${\bf q} = \Pi_{\rm m} c_1 \zeta_{\rm m} (\Delta T/L_z) {\bf \hat{z}}$, is absorbed into the definition of $\kappa_{\rm F}$ as $\kappa_{\rm F}- \Pi_{\rm m} c_1 \zeta_{\rm m} \to \kappa_{\rm F} $. \par
Now we can calculate the thermal power input $Q_{\rm h}$ from the hot reservoir. We assume that Domenicali's equation [Eq.~(\ref{Eq:Domenicali01})] approximately holds for P, neglecting spin Joule heating~\cite{Tulapurkar11}. Using heat current conservation at the P/F interface, we have $q_{\rm h} = \kappa_{\rm F} (\Delta T/L_z) - |\delta {\bf q}|- \rho_{\rm P} j^2$, where $\rho_{\rm P}= 1/\sigma_{\rm P}$. From this, we have 
%%%
\begin{eqnarray}
  Q_{\rm h} &=& K_{\rm F} \Delta T + c_2 \Pi_{\rm m} T (l_y/l_z) I_y - R I_y^2 
\end{eqnarray}
%%%
for $Q_{\rm h}= (l_x l_y) q_{\rm h}$. At the same time, the output power is given by 
%%%
\begin{eqnarray}
  P_{\rm out} &=& R_{\rm load} I_y^2, 
\end{eqnarray}
%%%
so that after a lengthy but straightforward calculation, the efficiency $\eta_{\rm SSE}=P_{\rm out}/Q_{\rm h}$ is obtained: 
%%%
\begin{eqnarray}
  \eta_{\rm SSE}(r) &=& \eta_{\rm C} \, g(r), 
\end{eqnarray}
%%%
where the characteristic function $g(r)$ is defined by 
%%%\kappa_{\rm F}
\begin{eqnarray}
  g(r) &=& 
  \frac{r}{(\frac{L_z}{l_z})\frac{(1+r)^2 }{Z_{\rm SSE} T_{\rm h}}
    - (\frac{L_z}{l_z}) \xi(1+r)- \frac{\Delta T}{T_{\rm h}} }, \\
  \xi &=& (c_2/c_1) (\sigma_{\rm P}/\sigma_{\rm m}), 
  \label{Eq:xi}
\end{eqnarray}
%%%
and the figure of merit of the LSSE device $Z_{\rm SSE}$ is given by 
%%%
\begin{equation}
  Z_{\rm SSE} = \frac{(\theta_{\rm SH} \alpha_{\rm S})^2}{\kappa_{\rm F} \rho_{\rm P}}. 
  \label{Eq:Z_S01}
\end{equation}
%%%
By maximizing $\eta_{\rm SSE}$ with respect to $r$, the optimized efficiency in the limit of $\Delta T \ll T_{\rm h}$ is calculated to be 
%%%
\begin{eqnarray}
  \eta_{\rm SSE}^* &=& \eta_{\rm C} \left( \frac{l_z}{L_z} \right) 
  \frac{1}{\xi} \left( \frac{1-r^*}{1+r^*}\right),  
  \label{Eq:etamax_SSE} \\
  r^* &=& \sqrt{1- \xi Z_{\rm SSE} T_{\rm h} }, 
\end{eqnarray}
%%%
where the functional form again has a characteristic of transverse devices. Note that, because the result is quite different from that of \cite{SSE_Cahaya2014APL} and several important steps to arrive at the final result therein are missing, we must conclude that it is almost impossible for us to draw parallel between the two approaches. In Table \ref{tab:1}, we summarize parameters that influence the efficiency of the LSSE thermoelectric generators. \par
\begin{table*}[tb]
  \caption{Parameters that influence the efficiency of the LSSE thermoelectric generators. } \label{tab:1}
  \begin{tabular}{l l}
    \hline \hline 
    $\theta_{\rm SH}$~~ & Spin Hall angle of P\\
    $\alpha_{\rm S}$~~ & Spin Seebeck coefficient [Eq. (26)]\\
    $\kappa_{\rm F}$~~ & Total thermal conductivity of F \\
    $\rho_{\rm P}$~~ & Electrical resistivity of P \\
    $l_z$~~ & Thickness of P\\
    $L_z$~~ & Thickness of F\\ 
    $\sigma_{\rm P}$~~ & Electrical conductivity of P, $\sigma_{\rm P}= 1/\rho_{\rm P}$ [Eq. (20)]\\
    $\sigma_{\rm m}$~~ & Magnon conductivity of F [Eq. (23)]\\ 
    $c_1$~~ & Ratio of spin current injected into P to thermal drift magnon current in F [Eq. (25)]\\
    $c_2$~~ & Ratio of spin current injected back into F to spin-Hall drift current in P [Eq. (28)]\\
    $\lambda$~~ & Spin diffusion length of P\\ 
    $\Lambda$~~ & Magnon diffusion length of F\\ 
    \hline \hline 
  \end{tabular}
\end{table*}
\subsection{Discussion}
First of all, it is important to note that, in comparison with the ANE device [Eq.~(\ref{Eq:ZANE})], the figure of merit of the LSSE device is determined by the thermal conductivity $\kappa_{\rm F}$ and the electrical resistivity $\rho_{\rm P}$ that are defined in two different materials, reflecting the fact that the charge current flows only through P whereas the thermal resistance is mainly given by F. Therefore, in contrast to $Z'_{\rm ANE}$ of the ANE device, $Z_{\rm SSE}$ of the LSSE device is free from the limitation of the Wiedemann-Franz law. \par
The optimized efficiency $\eta_{\rm SSE}^*$ is determined by three factors: the figure of merit $Z_{\rm SSE}$, the spin converting parameter $\xi$, and the geometrical factor $l_z/L_z$. Thus, for the calculation of $\eta_{\rm SSE}^*$, we need to estimate the magnitude of $\xi$ [Eq.~(\ref{Eq:xi})], in addition to the $Z_{\rm SSE}$ value. The parameter $\xi$ may be evaluated in the following way. First, since both the parameters $c_1$ and $c_2$ measure the ratio of a spin injection current to a bulk spin current, we could set $c_2/c_1 \approx 1$ for a rough estimate (a microscopic determination of the factor $c_2/c_1$ is beyond our scope since the present discussion is devoted to a phenomenological description of the efficiency). Next, the conductivity ratio may be estimated using Einstein relations, $\sigma_{\rm P}= e^2 D_{\rm P} g_{\rm P}(\epsilon_{\rm F})$ and $\sigma_{\rm m}= e^2 D_{\rm m} n_{\rm m}/ (k_{\rm B} T)$, where $D_{\rm P}$ and $D_{\rm m}$ are respectively the electron diffusion coefficient of P and magnon diffusion coefficient of F, $g_{\rm P}(\epsilon_{\rm F})$ is the electron density of states per unit volume at the Fermi energy, and $n_{\rm m}$ is the magnon number density. With these estimates, we obtain 
%%%
\begin{equation}
  \xi \approx  \left( \frac{D_{\rm P}}{D_{\rm m}} \right) \frac{g_{\rm P}(\varepsilon_{\rm F})}{n_{\rm m}/(k_{\rm B}T)}, 
\end{equation}
%%%
where $n_{\rm m}$ can be calculated by $n_{\rm m} = \int d \epsilon g_{\rm m} (\epsilon) f_{\rm BE}(\epsilon)$ with the magnon density of states per unit volume $g_{\rm m}(\epsilon)$ and the Bose-Einstein distribution function $f_{\rm BE}(\epsilon)$. \par
Now we discuss the upper limit of the efficiency $\eta_{\rm SSE}^*$. Similarly to the isothermal figure of merit of the ANE device [Eq.~(\ref{eq:ZTrange})], the isothermal LSSE figure of merit is expected to satisfy $0 \le Z_{\rm SSE} T \le 1$. Then, the maximum allowed efficiency of the LSSE device [Eq.~(\ref{Eq:etamax_SSE})] is achieved for $r^*=0$ (i.e., $\xi Z_{\rm SSE}T_{\rm h}=1$), which yields 
%%%
\begin{equation}
\eta_{\rm SSE}^{\rm max} = \eta_{\rm C} \left( \frac{l_z}{L_z} \right) \frac{1}{\xi}. 
\label{Eq:etaSSEmax}
\end{equation}
%%%
Note that the existence of such a maximum allowed efficiency requires a condition that $0 \le 1/\xi \le 1$; otherwise the situation $r^*=0$ cannot be realized. Note also that, although at first glance it appears that a use of a thin magnetic layer ($L_z \ll l_z$) would result in a very large efficiency, this naive expectation does not work since $1/\xi$ is proportional to $L_z/\Lambda$ for $L_z/\Lambda \ll  1$. Anyway, as is obvious from the structure of the present calculation, the efficiency of the LSSE device is bound by the constraint of the second law of thermodynamics. Therefore, there must be a condition $(l_z/L_z)(1/\xi) \leq 1$ to ensure that the upper limit of the maximum allowed efficiency is the Carnot efficiency, a microscopic derivation of which is left to future studies. However, the efficiency of present LSSE devices is still very low; making use of typical values for Pt/YIG systems: $l_z = 10~\textrm{nm}$, $L_z = 1~\mu \textrm{m}$, $\rho_{\rm P} = 0.1~\mu \Omega \textrm{m}$, $\kappa_{\rm F} = 7~\textrm{W/mK}$, $\theta_{\rm SH} \alpha_{\rm S} = 1~\mu \textrm{V/K}$, and $\xi = 1$, the figure of merit and the ratio of the optimized efficiency to the Carnot efficiency are respectively estimated to be $Z_{\rm SSE} T = 4 \times 10^{-4}$ and $\eta_{\rm SSE}^* / \eta_{\rm C} = 1 \times 10^{-4}~\%$ at $T = 300~\textrm{K}$, where $\eta_{\rm SSE}^*$ has a very little dependence on $\xi$ when $\xi Z_{\rm SSE} T \ll 1$. Therefore, the dramatic enhancement of the LSSE thermopower and the reduction of thermal conductivity of F are necessary for realistic thermoelectric applications of the LSSE. The determination of the optimum geometrical factor, $l_z/L_z$, is also important for improving the efficiency, while the increase of  $l_z/L_z$ may conflict with the thickness dependence of the LSSE because the LSSE thermopower usually decreases with increasing $l_z$ \cite{SSE_Qu2013PRL,SSE_Kikkawa2013PRB,SSE_Qu2014PRB,SSE_Saiga2014APEX,SSE_Qu2015PRB_Cr} and decreasing $L_z$ \cite{SSE_Kirihara2012NatMat,SSE_Kikkawa2015PRB,SSE_Kehlberger2015PRL,SSE_Ritzmann2015,SSE_Guo2015}. \par
%
%
%%%%%%%%%%%%%%%%%%%%%%%%%%%%%%%%%%%%%%%%%%%%%%%%%%%%
\section{Demonstrations for thermoelectric applications} \label{sec:application}
%%%%%%%%%%%%%%%%%%%%%%%%%%%%%%%%%%%%%%%%%%%%%%%%%%%%
%
%
In this section, we introduce several approaches for future thermoelectric applications of the LSSE devices, which include the demonstration of the LSSE-based thermoelectric generation in coating films, flexible devices, all-ferromagnetic devices, and multilayer films. The former two structures are under development to exploit the versatility and scalability of the LSSE devices, while the latter two are to improve their thermoelectric performance. \par
\subsection{Spin thermoelectric coating} \label{sec:coating}
The concept of a coating-based LSSE device called ``spin thermoelectric (STE) coating'' was first proposed in 2012 \cite{SSE_Kirihara2012NatMat}. The STE-coating device consists of a metal/magnetic insulator bilayer film directly coated on a heat source. Thanks to the simple structure and straightforward scaling law of the LSSE, the STE coating is potentially applicable to the implementation of large-area thermoelectric devices onto various-shaped heat sources, such as the surfaces of electronic instruments and automobile bodies. \par
\begin{figure}[tb]
\begin{center}
\includegraphics{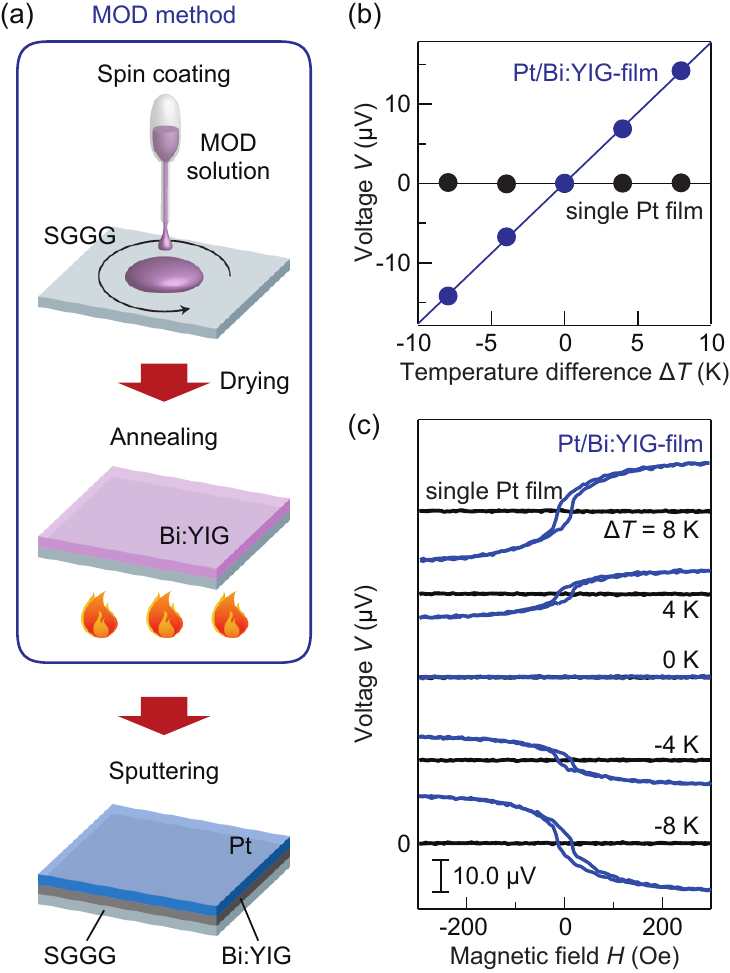}
\caption{(a) Preparation process for the Pt/Bi:YIG STE-coating device. (b) $\Delta T$ dependence of $V$ in the Pt/Bi:YIG-film sample and single Pt film. (c) $H$ dependence of $V$ in the Pt/Bi:YIG-film sample and single Pt film for various values of $\Delta T$. All the measurements were performed at room temperature. }\label{fig:STEcoating}
\end{center}
\end{figure}

The STE coating was demonstrated by using a spin-coating method, which may realize large-area thermoelectric devices in a highly productive way. Here, we used Bi:YIG and Pt as the magnetic insulator and metallic film layers, respectively. The STE-coating films were formed by using the simple fabrication steps illustrated in Fig. \ref{fig:STEcoating}(a). First, a Bi:YIG film was formed on a (GdCa)$_3$(GaMgZr)$_5$O$_{12}$ (substituted gadolinium gallium garnet: SGGG) substrate, which has good lattice-matching properties with Bi:YIG, by means of a metal-organic decomposition (MOD) method \cite{SSE_Kirihara2012NatMat}. The MOD method consists of simple three step  processes: spin coating, drying, and annealing of the MOD solution containing the constituent elements, which enables the fabrication of thin magnetic insulator films without using specialized and costly equipment, such as vacuum deposition apparatus. Subsequently, the Pt film was formed by sputtering over the whole surface of the Bi:YIG film. The thicknesses of the Bi:YIG and Pt films are 120 nm and 10 nm, respectively. Importantly, although the surface area of the samples used here is small ($8 \times 2~\textrm{mm}^2$), the above process can be easily applied to large-area manufacturing, since it does not require patterning steps such as photolithography and electron-beam lithography. \par
In Figs. \ref{fig:STEcoating}(b) and \ref{fig:STEcoating}(c), we respectively show the $\Delta T$ and $H$ dependences of $V$ in the Pt/Bi:YIG STE-coating film. The clear LSSE voltage was found to appear even in this thin film structure prepared by the MOD coating method. We also checked that no signal appears in a plain Pt film sputtered directly on a SGGG substrate, indicating that the thin Bi:YIG film works as a thermoelectric generator. We found that a crystalline Bi:YIG film can be grown even onto a glass substrate by using the MOD method \cite{SSE_Kirihara2012NatMat}, and the STE coating is applicable even onto amorphous surfaces. Such versatile implementation of thermoelectric functions may open opportunities for various applications making full use of omnipresent heat. \par
\subsection{Flexible spin Seebeck devices} \label{sec:flexible}
Most of the conventional thermoelectric devices are rigid, and not easily applicable onto curved or uneven heat sources. Because the conventional thermoelectric device consists of a number of $\Pi$-structured thermocouples connected electrically in series, it is vulnerable to bending stresses, making it difficult to construct flexible devices. In contrast, the LSSE device has a high affinity for flexible thermoelectric generation owing to its simple structure and scaling law; it will be applicable on various heat sources, paving the way to versatile thermoelectric generators or sensors. \par
Recently, we demonstrated the construction of a LSSE-based flexible thermoelectric sheet \cite{SSE_Kirihara2015}. To construct the flexible LSSE device, we used spray-coating technique called ``ferrite plating'' \cite{Abe1992,Kondo2007}, which enables the fabrication of ferrimagnetic ferrite thin films. Since conventional ferrite-film preparation techniques, including sputtering \cite{SSE_Niizeki2015AIPAdv_CFO}, liquid phase epitaxy \cite{Qiu_spin-pump_2013APL}, and pulsed laser deposition \cite{spin mixing concept}, require high temperature processes (ranging from 400 $^\circ$C to 800 $^\circ$C) for crystallizing ferrites, they cannot be used for the formation of ferrite films on heat-labile soft materials, such as plastics. By contrast, the ferrite plating method is based on chemical reaction processes, and thus does not need any high temperature processes, enabling the coating of ferrite films on a variety of substrates including flexible plastic films. \par
\begin{figure}[tb]
\begin{center}
\includegraphics{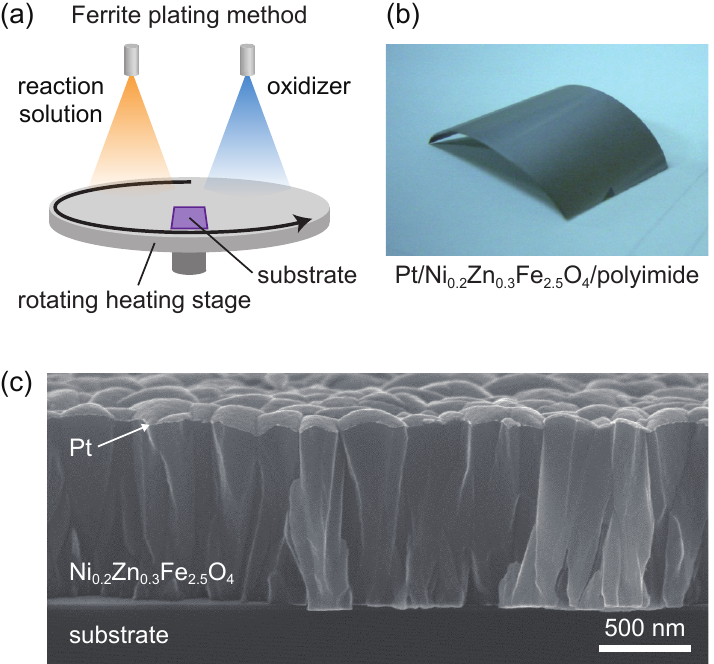}
\caption{(a) A schematic illustration of the ferrite-plating method. (b) A photograph of a LSSE-based flexible thermoelectric sheet, where a Pt/Ni$_{0.2}$Zn$_{0.3}$Fe$_{2.5}$O$_{4}$ film was formed on a 25-$\mu$m-thick polyimide flexible substrate. (c) A scanning electron microscope image of a Ni$_{0.2}$Zn$_{0.3}$Fe$_{2.5}$O$_{4}$ film grown on a thermally oxidized silicon substrate by the ferrite-plating method. }\label{fig:Flexible}
\end{center}
\end{figure}
The fabrication process of the ferrite plating method is very simple, which also requires no specialized and costly equipment. As schematically illustrated in Fig. \ref{fig:Flexible}(a), the ferrite film can be grown simply by spraying an aqueous reaction solution and an oxidizer simultaneously onto a substrate, mounted on a rotating heating stage \cite{Abe1992,Kondo2007}. The thickness of the ferrite film can be controlled by the spray volume and time and can be dependent on surface conditions of substrates. Importantly, all the processes of the ferrite plating method can be performed below 100 $^\circ$C, applicable to heat-labile plastic substrates. \par
By using the ferrite plating method, we successfully fabricated the flexible thermoelectric sheet based on the LSSE [see a photograph in Fig. \ref{fig:Flexible}(b)]. The flexible LSSE device consists of a ferrimagnetic Ni$_{0.2}$Zn$_{0.3}$Fe$_{2.5}$O$_{4}$ thin film, grown on a flexible polyimide substrate by the ferrite plating, and a Pt film, sputtered on the Ni$_{0.2}$Zn$_{0.3}$Fe$_{2.5}$O$_{4}$ film. As shown in Fig. \ref{fig:Flexible}(b), the Pt/Ni$_{0.2}$Zn$_{0.3}$Fe$_{2.5}$O$_{4}$/polyimide sheet is highly flexible and bendable. We observed clear LSSE signals in the flexible thermoelectric sheet, where the magnitude of the LSSE signals is comparable to that in the conventional rigid LSSE devices \cite{SSE_Kirihara2015}. We also demonstrated that the ferrite plating method is applicable to not only plastic substrates but also glass substrates; it is useful for constructing low-cost LSSE devices. \par
A noticeable feature of the ferrite film grown by the ferrite plating method is its columnar grain structure [see the cross-sectional scanning electron microscope image of the Ni$_{0.2}$Zn$_{0.3}$Fe$_{2.5}$O$_{4}$ film in Fig. \ref{fig:Flexible}(c)]. Here, the crystal orientation of Ni$_{0.2}$Zn$_{0.3}$Fe$_{2.5}$O$_{4}$ is coherently aligned within each columnar grain \cite{SSE_Kirihara2015}. Such columnar structure can be suitable for the flexible LSSE devices because of the following two reasons. First, in the LSSE configuration, the spin current is generated along the columnar structure, and thus less affected by grain scattering. Second, the columnar grain boundary can work as a stress-relieving cushion when the film is bent, leading to the high flexibility and bending tolerance. \par
The ferrite plating method potentially enables the direct coating of thermoelectric functions onto various surfaces over a large area, expanding the versatility and utility of the LSSE devices. Significantly, the flexible thermoelectric sheet based on the LSSE, demonstrated here, has remarkably low thermal resistance because of the thick-substrate-free structure. Therefore, it may be suitable for heat-flow sensing applications \cite{SSE_Kirihara2015}. \par
\subsection{Hybrid thermoelectric generation based on spin Seebeck and anomalous Nernst effects} \label{sec:ferro}
\begin{figure*}[tb]
\begin{center}
\includegraphics{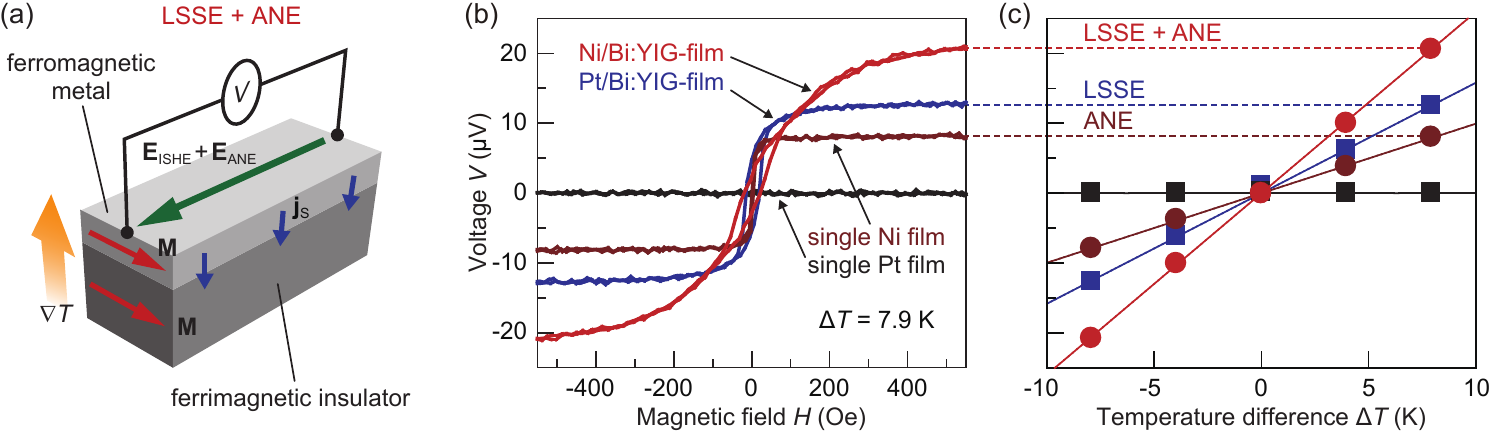}
\caption{(a) A schematic illustration of the hybrid thermoelectric generation based on the LSSE and ANE in a ferromagnetic metal/ferrimagnetic insulator junction system. (b) $H$ dependence of $V$ in the Ni/Bi:YIG-film sample, Pt/Bi:YIG-film sample, single Ni film, and single Pt film at $\Delta T = 7.9~\textrm{K}$. (c) $\Delta T$ dependence of $V$ in the Ni/Bi:YIG-film sample, Pt/Bi:YIG-film sample, single Ni film, and single Pt film. The thicknesses of the Ni, Pt, and Bi:YIG films are 10 nm, 10 nm, and 120 nm, respectively. All the measurements were performed at room temperature. }\label{fig:ANEhybrid}
\end{center}
\end{figure*}
As shown in Fig. \ref{fig:LSSEmaterials}, most of the LSSE experiments to date have been performed by using Pt as a metallic layer. Pt is useful for investigating the physics of the LSSE, since it enables sensitive detection of the thermally generated spin currents and the spin-current-related parameters, such as the spin diffusion length and spin Hall angle, of Pt have been well studied \cite{ISHE_Sinova}. However, Pt is unsuitable for thermoelectric applications of the LSSE because of its high material cost. Furthermore, even the spin-charge conversion efficiency of Pt is insufficient for obtaining adequate thermoelectric performance. \par
Recently, the ISHE has been investigated not only in paramagnetic materials but also in ferromagnetic materials by means of the LSSE experiments \cite{SSE_Miao2013PRL_PyYIG,SSE_Kikkawa2013PRB,SSE_Azevedo2014APL_PyYIG,SSE_Wu2014APL_Fe3O4,SSE_Tian2015APL,SSE_TSeki2015APL} (Fig. \ref{fig:LSSEmaterials}). The main purpose of these studies is microscopic understanding of the ISHE in ferromagnets. In contrast, we focus on ferromagnetic materials as low-cost replacements for Pt in the LSSE devices. Here, as a simple model case, we consider a ferromagnetic metal/ferrimagnetic insulator junction system under a temperature gradient. In this system, the ANE is induced in the ferromagnetic metal layer in a conventional manner. In addition to the ANE contribution, if the spin current is injected into the ferromagnetic metal from the ferrimagnetic insulator via the LSSE and the ISHE appears in the ferromagnetic metal, the hybrid thermoelectric generation based on the combination of the LSSE and ANE is realized [Fig. \ref{fig:ANEhybrid}(a)]. The previous studies showed that the direction of ${\bf E}_{\rm ISHE}$ is the same as that of ${\bf E}_{\rm ANE}$ in various ferromagnetic materials \cite{SSE_Miao2013PRL_PyYIG,SSE_Kikkawa2013PRB,SSE_Azevedo2014APL_PyYIG,SSE_Wu2014APL_Fe3O4,SSE_Tian2015APL,SSE_TSeki2015APL}. Even when the spin Hall angle of the ferromagnetic metal is smaller than that of Pt, the shortfall of the ISHE voltage can be compensated by the superposition of the ANE voltage. In fact, the thermoelectric output of all-ferromagnetic devices can even be better than that of the conventional Pt-based devices, as demonstrated below. \par
In Figs. \ref{fig:ANEhybrid}(b) and \ref{fig:ANEhybrid}(c), we compare the transverse thermoelectric voltage between ferromagnetic Ni/Bi:YIG-film and conventional Pt/Bi:YIG-film systems. The Bi:YIG layer of the Ni/Bi:YIG-film and Pt/Bi:YIG-film samples were formed on a SGGG substrate by means of the MOD method. We found that the Ni/Bi:YIG-film sample exhibits the clear $V$ signal of which the magnitude is more than twice greater than that of the ANE voltage in a plain Ni film directly placed on the substrate, indicating that the signal in the Ni/Bi:YIG-film sample is attributed to the superposition of the ANE in the Ni layer and the ISHE induced by the spin current injected from the Bi:YIG layer. Importantly, the magnitude of $V$ in the Ni/Bi:YIG-film sample is even greater than that in the conventional Pt/Bi:YIG-film sample [Figs. \ref{fig:ANEhybrid}(b) and \ref{fig:ANEhybrid}(c)]. Since the electrical resistivity of Ni is comparable to or smaller than that of Pt, the Ni/Bi:YIG-film sample shows the better thermoelectric performance than the Pt/Bi:YIG-film sample in terms of not only the output voltage but also the output current or power. This result suggests that the hybrid thermoelectric generation using ferromagnetic metal/ferrimagnetic insulator junction systems will be useful for constructing noble-metal-free, low-cost, and efficient LSSE devices. \par
\subsection{Spin Seebeck effect in multilayer devices} \label{sec:multilayer}
\begin{figure}[tb]
\begin{center}
\includegraphics{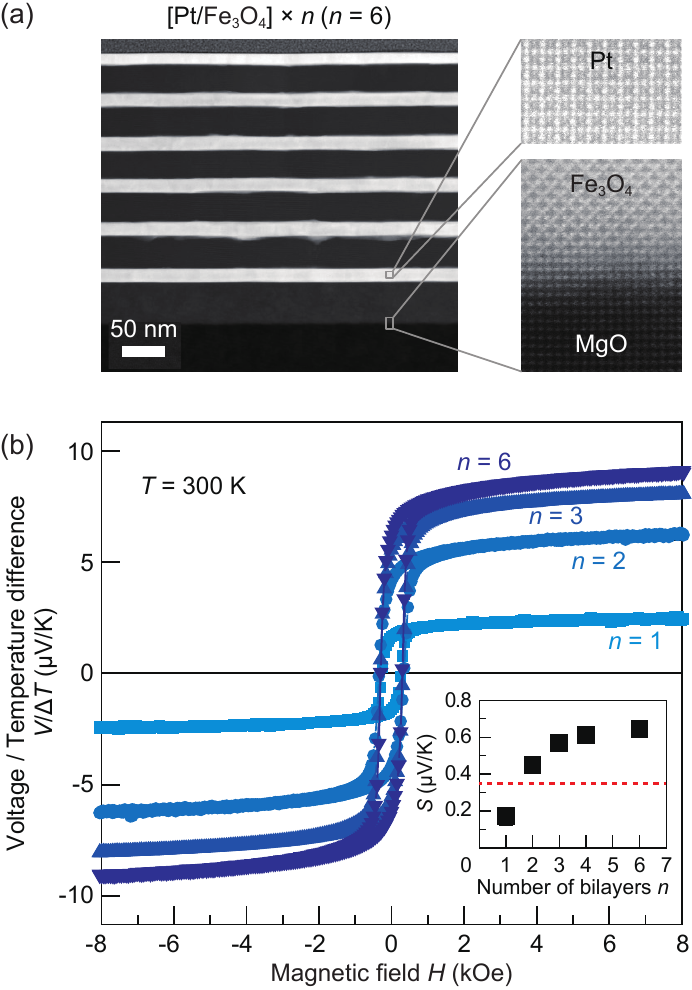}
\caption{(a) Scanning transmission electron microscope image of the cross section of the [Pt/Fe$_3$O$_4$] $\times$ 6 system, where the thickness of the Pt (Fe$_3$O$_4$) layers is 17 nm (34 nm). The magnified views of the Pt layer and Fe$_3$O$_4$/MgO interface show the high crystalline quality of the multilayers. (b) $H$ dependence of $V/\Delta T$ in the [Pt/Fe$_3$O$_4$] $\times n$ systems for various values of the Pt/Fe$_3$O$_4$-bilayer number $n$ at $T = 300~\textrm{K}$. The inset to (b) shows the $n$ dependence of $S$ in the [Pt/Fe$_3$O$_4$] $\times n$ systems. The red dotted line represents the upper limit of the LSSE voltage for the parallel circuit of connected Pt/Fe$_3$O$_4$ bilayers. }\label{fig:multilayer-exp}
\end{center}
\end{figure}
\begin{figure*}[tb]
\begin{center}
\includegraphics{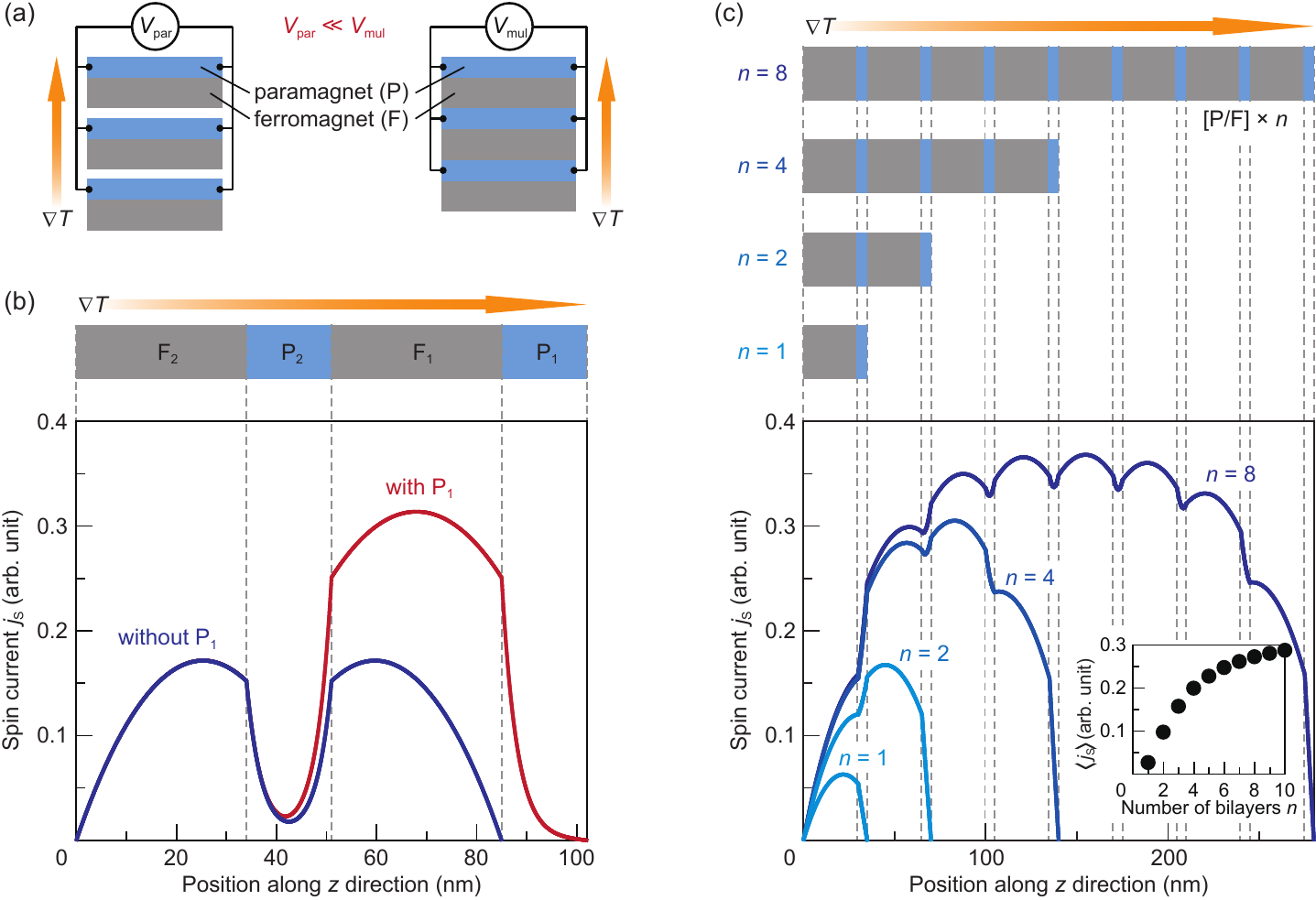}
\caption{(a) Comparison of the LSSE voltage between paramagnet (P)/ferromagnet (F) bilayers electrically connected in parallel and an alternately-stacked P/F multilayer film. The LSSE voltage in the latter $V_{\rm mul}$ was found to be much greater than that in the former $V_{\rm par}$. (b) Comparison of the spin-current $j_{\rm S}$ profiles between P$_1$/F$_1$/P$_2$/F$_2$ and F$_1$/P$_2$/F$_2$ systems. (c) $j_{\rm S}$ profiles calculated for the [P/F] $\times$ $n$ multilayer systems for various values of the P/F-bilayer number $n$. The inset to (c) shows the $n$ dependence of the spin-current magnitude averaged over all the P layers $\langle j_{\rm S} \rangle$. The discussion on these calculation results is detailed in \cite{SSE_ML_Ramos}. }\label{fig:multilayer1}
\end{center}
\end{figure*}

To realize efficient thermal spin-current generation, the LSSE has recently been investigated in multilayer systems comprising alternately-stacked paramagnet (P)/ferromagnet (F) films \cite{SSE_ML_Lee,SSE_LSMO_La2NiMnO6,SSE_ML_Ramos}. The recent studies have revealed that the LSSE voltage in [P/F] $\times$ $n$ systems significantly and monotonically increases with increasing the number of the P/F bilayers $n$. For example, in \cite{SSE_ML_Ramos}, the magnitude of the LSSE voltage in [Pt/Fe$_3$O$_4$] $\times$ 6 systems was observed to be enhanced by a factor of 4--6 compared with that in [Pt/Fe$_3$O$_4$] $\times$ 1 bilayer systems (Fig. \ref{fig:multilayer-exp}). Since this LSSE-voltage enhancement is unaccompanied by the increase of the internal resistance, the output power also increases with increasing $n$, a situation different from the case of the LSSE-voltage enhancement by the spin Hall thermopile \cite{SSE-Uchida2014JPCM,SSE_Uchida2012APEX}. The observed $n$ dependence of the LSSE voltage in the [P/F] $\times n$ multilayer systems is beyond conventional expectations based on the situation that the systems are merely regarded as several independent P/F bilayers electrically connected in parallel, where the output {\it voltage} is not enhanced while the output {\it power} is enhanced owing to the reduction of the internal resistance [Fig. \ref{fig:multilayer1}(a)] \cite{SSE_ML_Ramos}. Importantly, this LSSE-voltage enhancement cannot be explained even when the spin-current injection into P from both the top and bottom F layers is taken into account, where the upper limit of the LSSE enhancement is twice of the voltage in the single P/F bilayer; as shown in Fig. \ref{fig:multilayer-exp}(b), the observed enhancement is much greater than this upper limit. \par
Our current interpretation of the mechanism of the LSSE enhancement in the P/F multilayer systems is summarized as follows \cite{SSE_ML_Ramos}. The essence of the LSSE enhancement is the boundary conditions for spin currents flowing normal to the P/F interfaces, which affect the magnitude and spatial profile of the spin currents generated by the LSSE. Here we assume the following two boundary conditions: (i) spin currents must disappear at the top and bottom surfaces of the multilayer systems and (ii) spin currents in the P and F layers are continuous at the interfaces. Although spin currents in paramagnetic metals and ferrimagnetic insulators are respectively carried by conduction electrons and spin waves, the boundary condition (ii) allows us to treat these spin currents in the same manner in the following phenomenological discussions. Let us now compare a spin-current profile in a P$_1$/F$_1$/P$_2$/F$_2$ system with that in a F$_1$/P$_2$/F$_2$ system without the top P layer (P$_1$) [Fig. \ref{fig:multilayer1}(b)], where P$_{1}$ and P$_{2}$ are good spin sinks. According to the boundary condition (i), the spin current is eliminated at the top of the F$_1$ layer in the F$_1$/P$_2$/F$_2$ system. However, this is not the case for the P$_1$/F$_1$/P$_2$/F$_2$ system; the spin current remains a large value at the P$_1$/F$_1$ interface, while it must disappear at the top of the P$_1$ layer. As shown in Fig. \ref{fig:multilayer1}(b), this difference results in the enhancement of the spin currents near the P$_1$/F$_1$ interface, a situation consistent with the prediction in \cite{SSE_Hoffman2013PRB}. By applying the above discussion, we calculated out-of-plane spin-current profiles in the [P/F] $\times$ $n$ systems for various values of $n$ [Fig. \ref{fig:multilayer1}(c)] and the $n$ dependence of the spin-current magnitude averaged over all the P layers $\langle j_{\rm S} \rangle$ [the inset to Fig. \ref{fig:multilayer1}(c)] (note that $\langle j_{\rm S} \rangle$ can be regarded as an observable quantity in the measurements of the LSSE in the P/F multilayer systems, since the P layers are electrically connected with each other due to the small thickness and resultant small out-of-plane electrical resistance of the F layers \cite{SSE_ML_Ramos}). Importantly, the magnitude of $\langle j_{\rm S} \rangle$ monotonically increases with increasing $n$. The physics behind is that, thanks to the multilayer structure, the spin current in the P interlayers acquires a new length scale and boundary value. This phenomenological interpretation is consistent with the experimental results; similar $n$ dependence of the LSSE voltage was observed in the [Pt/Fe$_3$O$_4$] $\times$ $n$ systems [compare the inset to Fig. \ref{fig:multilayer-exp}(b) with that to Fig. \ref{fig:multilayer1}(c)] \cite{SSE_ML_Ramos}. The [Pt/Fe$_3$O$_4$] $\times$ 6 system holds the current record of the LSSE thermopower at room temperature (Fig. \ref{fig:trend}). Since the enhancement of the LSSE based on this mechanism strongly depends on the spin diffusion length (magnon diffusion length) of the P (F) layer, the determination of optimum thicknesses of each layer and optimum P/F material combination is crucial for further improvement of the thermoelectric performance of the LSSE devices. \par
%
%
%
%%%%%%%%%%%%%%%%%%%%%%%%%%%%%%%%%%%%%%%%%%%%%%%%%%%%
\section{Conclusions and prospects} \label{sec:conclusion}
%%%%%%%%%%%%%%%%%%%%%%%%%%%%%%%%%%%%%%%%%%%%%%%%%%%%
%
%
\begin{figure*}[htb]
\begin{center}
\includegraphics{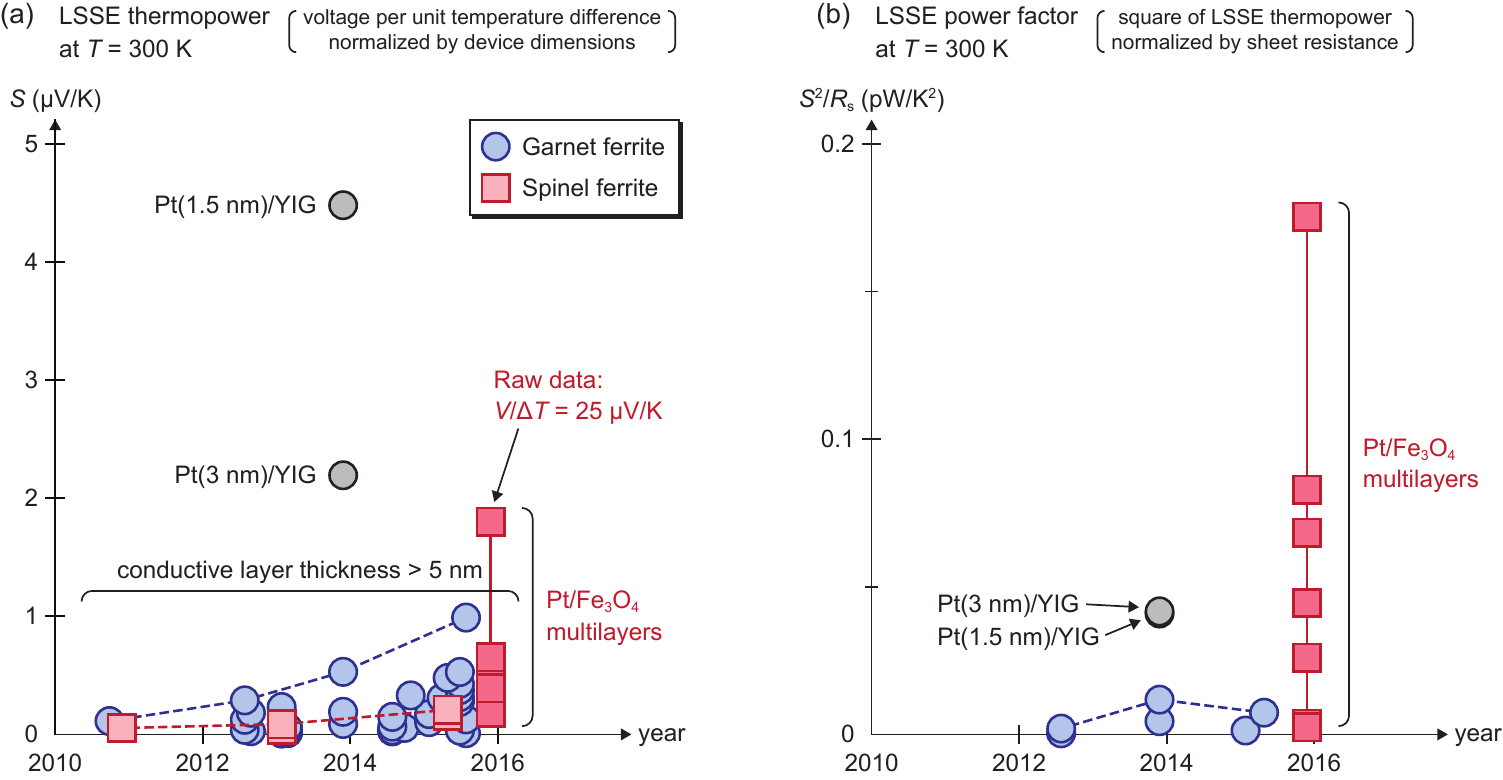}
\caption{Time-series trends of the LSSE thermopower $S$ (a) and the LSSE power factor $S^2/R_{\rm s}$ (b) at $T = 300~\textrm{K}$ in the typical LSSE devices \cite{SSE_Uchida2010APL_1,SSE_Kirihara2012NatMat,SSE-Uchida2014JPCM,SSE_MnZnFO,SSE_Uchida2012APEX,SSE_Kikkawa2013PRL,SSE_Meier2013PRB_NFO,SSE_Ramos2013APL_Fe3O4,SSE_Uchida2013PRB,SSE_Kikkawa2013PRB,SSE_Lustikova2014JAP,SSE_Uchida2014PRX,SSE_BiYIG_Kikuchi,SSE_Qiu2015JPD,SSE_Niizeki2015AIPAdv_CFO,SSE_Uchida2015PRB_PMA,SSE_Qiu2015APEX,SSE_Kikkawa2015PRB,SSE_ML_Ramos,ANE_Uchida2015}. Although the output voltage can be increased simply by reducing the conductive layer thickness of the LSSE devices, such thin films cannot exhibit high output power due to their high internal resistance (see gray circle data points). For fair comparison, the experimental data for the LSSE devices with the spin Hall thermopiles are not shown in these trends. For example, in \cite{SSE-Uchida2014JPCM}, the large LSSE voltage of $V/\Delta T \sim 0.4~\textrm{mV/K}$ was realized by using the Pt-Au thermopile structure, although its internal resistance is very high. }\label{fig:trend}
\end{center}
\end{figure*}
In this article, we reviewed the experimental results of the LSSE in insulator-based systems from the viewpoint of thermoelectric applications. The basic structure of the LSSE device is a simple bilayer film comprising a magnetic insulator and a metallic or other conductive material; an out-of-plane temperature gradient in the magnetic insulator layer induces an in-plane electric field in the metallic layer via spin-current generation across the interface. Owing to this configuration, the LSSE device exhibits the following unique characteristics. Firstly, the figure of merit of the LSSE device is free from the Wiedemann-Franz law, and can be optimized by selecting the combination of a magnetic insulator with low thermal conductivity and a metallic film with low electrical resistivity. Secondly, the LSSE device follows the convenient scaling law; the maximum extractable power of the LSSE device is proportional to the device area, and can be enhanced simply enlarging the device without constructing complicated modules. The simple structure and scaling law of the LSSE device make it possible to implement thermoelectric functions onto various heat sources by using versatile and low-cost fabrication processes. \par
The thermoelectric performance of the LSSE device cannot be simply compared with that of conventional thermoelectric devices based on the Seebeck effect because of the different device configurations and driving principles. The theory presented in Sec. \ref{sec:calc-efficiency} shows that the thermoelectric conversion efficiency of the LSSE device is characterized by the figure of merit $Z_{\rm SSE}T$, described by Eq. (\ref{Eq:Z_S01}). Here, $Z_{\rm SSE}T$ is defined as proportional to the square of the LSSE thermopower and inversely proportional to the thermal conductivity of a magnetic insulator and electrical resistivity of a metallic film, which is similar to the case of the conventional thermoelectric devices. However, the figure of merit of the LSSE device is defined in the range of $0 \leq Z_{\rm SSE}T \leq 1$, which is a characteristic of transverse thermoelectric devices. \par
Despite the potential advantages of the LSSE device, the current stage of the SSE research is still far from realistic thermoelectric applications. This is mainly because the LSSE thermopower is very small at present, although there is plenty of scope for the performance improvement. The LSSE thermopower can be enhanced by improving the thermal spin-current generation efficiency in the magnetic insulator, spin Hall angle in the metallic layer, and spin mixing conductance at the metal/insulator interface. In fact, owing to the efforts after the discovery of the LSSE, the thermoelectric performance of the LSSE device is being improved, especially, in recent years with the advent of multilayer systems as shown in the time-series trends of the LSSE thermopower $S$ [Fig. \ref{fig:trend}(a)] and the LSSE power factor [Fig. \ref{fig:trend}(b)] in typical LSSE devices. Here, we define the LSSE power factor as the output power normalized by the applied temperature gradient and by the device area: 
\begin{equation}
\frac{S^2}{R_{\rm s}} = \left({ \frac{V}{\Delta T} \frac{L_z}{L_y}}\right)^2 \frac{l_y}{R l_x} = \frac{V^2}{R} \frac{1}{\nabla T^2} \frac{1}{l_x l_y}, 
\label{Eq:LSSE-PF}
\end{equation}
where $R_{\rm s}$ ($=Rl_x/l_y$) is the sheet resistance of the metallic layer of the LSSE device and we assume $l_{x(y)} = L_{x(y)}$. Figure \ref{fig:trend}(b) emphasizes again that multilayer systems are useful for improving the output power of the LSSE device due to the combination of the voltage enhancement and internal resistance reduction. Nevertheless, since even the current record of the thermoelectric performance of the LSSE is still inadequate, continuous development of the LSSE-based thermoelectric technology is necessary. For realistic thermoelectric applications, the reduction of the thermal conductivity (electrical resistivity) of the insulator layer (metallic layer) for the improvement of  $Z_{\rm SSE}T$ and optimum thermal design for the stable and continuous operation of the LSSE device are also indispensable. \par
In the field of spintronics, in addition to the SSE, a variety of novel phenomena in which the interplay of spin and heat plays a crucial role were discovered \cite{spincaloritronics-Heremans,spincaloritronics-Bauer}. Some of the thermo-spin phenomena, such as spin-dependent Seebeck effects \cite{Slachter2010NatPhys,Breton2011Nature}, Seebeck effects in magnetic tunnel junctions \cite{Walter2011NatMater}, and magnon-drag effects \cite{Costache2012NatMat}, are also potentially applicable to thermoelectric generation in nano-structured spintronics devices. Furthermore, at low temperatures, giant thermopower was observed in transverse SSE devices comprising non-magnetic semiconductors \cite{Jaworski2012Nature}. Although the origin of this giant thermoelectric effect is different from the SSE discussed in this article, it represents a great potential of spin-based thermoelectric technologies. However, these thermo-spin phenomena appear only in conductors; the utilization of insulators is an exclusive feature of the SSE-based thermoelectric generation. We anticipate that this unique feature of the SSE will lead to various thermoelectric applications, including energy harvesters, thermometers, infrared sensors, position detectors \cite{SSE-Uchida2011JJAP}, and user-interface devices. \par
%
%
%%%%%%%%%%%%%%%%%%%%%%%%%%%%%%%%%%%%%%%%%%%%%%%%%%%%
\section*{ACKNOWLEDGMENTS}
%%%%%%%%%%%%%%%%%%%%%%%%%%%%%%%%%%%%%%%%%%%%%%%%%%%%
%
%
The authors thank M. H. Aguirre, M. Akimoto, P. A. Algarabel, A. Anad\ifmmode \acute{o}\else \'{o}\fi{}n, J. Barker, G. E. W. Bauer, S. R. Boona, C. L. Chien, S. Daimon, S. T. B. Goennenwein, J. P. Heremans, B. Hillebrands, T. Hioki, D. Hou, S. Y. Huang, M. R. Ibarra, J. Ieda, R. Iguchi, K. Ihara, Y. Iwasaki, H. Jin, X. F. Jin, D. Kikuchi, M. Kl\"aui, S. Kohmoto, K. Kondo, I. Lucas, T. Manako, A. Matsuba, A. Miura, L. Morell\ifmmode \acute{o}\else \'{o}\fi{}n, T. Murakami, R. C. Myers, Y. Nakamura, T. Niizeki, T. Nonaka, J. Ohe, Y. Ohnuma, Y. Oikawa, T. Ota, T. Oyake, Z. Qiu, R. Ramos, S. M. Rezende, K. Sato, T. Seki, J. Shiomi, Y. Shiomi, H. Someya, S. Takahashi, K. Takanashi, Y. Tserkovnyak, B. J. van Wees, J. Xiao, N. Yamamoto, and A. Yagmur for valuable discussions. \par
This work was supported by PRESTO ``Phase Interfaces for Highly Efficient Energy Utilization'' from JST, Japan, Grant-in-Aid for Scientific Research (A) (No. 15H02012), Grant-in-Aid for Challenging Exploratory Research (No. 26600067), Grant-in-Aid for Scientific Research on Innovative Area, ``Nano Spin Conversion Science'' (No. 26103005) from MEXT, Japan, and NEC Corporation. T.K. is supported by JSPS through a research fellowship for young scientists (No. 15J08026). 
\par~\\~\\
\noindent{\bf Profiles: }\par~\\
{\bf Ken-ichi Uchida} was born in Kanagawa Prefecture, Japan, in 1986. He received the B.Eng. and M.Sc. Eng. degrees from Keio University, Yokohama, Japan, in 2008 and 2009. respectively, and the Ph.D. degree from Tohoku University, Sendai, Japan, in 2012. \par
He was an Assistant Professor at the Institute for Materials Research, Tohoku University from 2012 to 2014. He has been an Associate Professor at the Institute for Materials Research, Tohoku University, since 2014, and a PRESTO Researcher at Japan Science and Technology Agency, since 2012. He has worked on spintronics and spin caloritronics. He has developed thermal, acoustic, and plasmonic methods of generating spin currents and realized thermoelectric generation using insulators. \par~\\
{\bf Hiroto Adachi} was born in Aichi Prefecture, Japan, in 1975. He received the B.Sc., M.Sc., and D.Sc. degrees from Kyoto University, Kyoto, Japan, in 1998, 2000, and 2003, respectively. \par
He was a Postdoctoral Research Assistant at Okayama University, Okayama, Japan, from 2003 to 2007, at the Swiss Federal Institute of Technology (ETH) Zurich, Zurich, Switzerland, from 2007 to 2009, at the Institute for Materials Research, Tohoku University, Sendai, Japan, from 2009 to 2010, and at the Advanced Science Research Center, Japan Atomic Energy Agency, Tokai, Japan, from 2010 to 2011. Since 2011, he has been a Senior PD Fellow at the Advanced Science Research Center, Japan Atomic Energy Agency. His research interests include spintronics and superconductivity. \par~\\
{\bf Takashi Kikkawa} was born in Kanagawa Prefecture, Japan, in 1991. He received the B.Sc. and M.Sc. degrees from Tohoku University, Sendai, Japan, in 2013 and 2015, respectively. He is currently working toward the Ph.D. degree in science at Tohoku University. \par
Since 2015, he has been supported by the Japan Society for the Promotion of Science through a research fellowship for young scientists. His research interest includes the investigation of thermo-spin effects in magnetic materials. \par~\\
{\bf Akihiro Kirihara} was born in Saga Prefecture, Japan, in 1979. He received the B.Eng. and M.Eng. degrees from the University of Tokyo, Tokyo, Japan, in 2002 and 2004, respectively. \par
He has belonged to NEC Corporation since 2004. He was a Visiting Researcher at Technische Universit$\ddot{\textrm{a}}$t Kaiserslautern, Kaiserslautern, Germany, from 2013 to 2014. Since 2014, he has also been a Research Member of the ERATO ``Spin Quantum Rectification'' project at Japan Science and Technology Agency. His research interests include spintronics, spin caloritronics, and thermoelectric conversion devices. \par~\\
{\bf Masahiko Ishida} was born in Fukushima Prefecture, Japan, in 1972. He received the B.Eng., M.Eng., and D.Eng. degrees from University of Tsukuba, Tsukuba, Japan, in 1995, 1997, and 1999, respectively. \par
After spending one year of postdoctoral research at University of Tsukuba, he joined Fundamental Research Laboratories, NEC Corporation in 2000. He was also a Visiting Researcher at the California Nanosystems Institute, University of California, Los Angeles, California, USA, from 2007 to 2008. Since 2014, he has been leading the Spin Thermoelectrics Team at Smart Energy Research Laboratories, NEC Corporation, and the spin Seebeck effect utilization group of the ERATO ``Spin Quantum Rectification'' project at Japan Science and Technology Agency. He has worked on various fields of electronic and sensing device fabrication. \par~\\
{\bf Shinichi Yorozu} was born in Ibaraki Prefecture, Japan, in 1965. He received the B.Eng. degree from Hokkaido University, Hokkaido, Japan, in 1988, and the M.Eng. and Ph.D. degrees from the University of Tokyo, Tokyo, Japan, in 1990 and 1993, respectively. \par
He joined Fundamental Research Laboratories, NEC Corporation in 1993. He was a Visiting Researcher at the State University of New York at Stony Brook, New York, USA, from 1997 to 1998. He was a Senior Manager of the Fundamental and Environmental Research Laboratories, NEC Corporation in 2005. Since 2015, he has been a Deputy General Manager of Smart Energy Research Laboratories, NEC Corporation. He has responsibility for the research area of nanotechnology and quantum technology, which include spintronics, nano-photonic devices, quantum information, and printed electronics. \par~\\
{\bf Sadamichi Maekawa} was born in Nara Prefecture, Japan, in 1946. He received the B.Sc. and M.Sc. degrees from Osaka University, Osaka, Japan, in 1969 and 1971, respectively, and the D.Sc. degree from Tohoku University, Sendai, Japan, in 1975. \par
He was a Research Associate from 1971 to 1982, and an Associate Professor from 1982 to 1988 at the Institute for Materials Research, Tohoku University, Sendai, Japan, a Professor in the Faculty of Engineering, Nagoya University, Nagoya, Japan, from 1988 to 1997, and a Professor at the Institute for Materials Research, Tohoku University from 1997 to 2010. Since 2010, he has been a Director of the Advanced Science Research Center, Japan Atomic Energy Agency, Tokai, Japan. His main research interests include theories of electronic properties in strongly correlated electron systems, in particular, high-temperature superconductors and orbital physics in transition metal oxides, and theories of transport phenomena in magnetic nanostructures. \par
Dr. Maekawa received Fellow of Institute of Physics (U.K.) in 1999, the Humboldt Award (Germany) in 2001, APS Fellow in 2007, the IUPAP Magnetism Award and Neel Medal in 2012, the Honoris Cause Doctorate of University of Zaragoza (Spain) in 2013, and several Japanese awards. \par~\\
{\bf Eiji Saitoh} was born in Tokyo, Japan, in 1971. He received the B.Eng., M.Eng., and Ph.D. degrees from the University of Tokyo, Tokyo, Japan, in 1996, 1998, and 2001, respectively. \par
He was a Research Associate in the Department of Physics, Keio University, Yokohama, Japan, from 2001 to 2006, and an Assistant Professor in the Department of Applied Physics and Physico-Informatics, Keio University, from 2006 to 2009. He has been a Professor at the Institute for Materials Research, Tohoku University, Sendai, Japan, since 2009, a Professor at the WPI Advanced Institute for Materials Research, Tohoku University, since 2012, and a Visiting Group Leader at the Advance Science Research Center, Japan Atomic Energy Agency, Tokai, Japan, since 2010. Since 2014, he has been a Research Director of the ERATO ``Spin Quantum Rectification'' project at Japan Science and Technology Agency. He was also a Visiting Scholar at the Cavendish Laboratory, University of Cambridge, Cambridge, U.K., in 2004 and a PRESTO Researcher at the Japan Science and Technology Agency, from 2007 to 2011. He has worked on various fields of condensed matter physics such as strongly correlated electron systems, nanomagnetics, and spintronics. His research is now focused on spin-related phenomena including the generation, detection, and manipulation of spin currents. \par
Dr. Saitoh received the IUPAP Young Scientist Award in 2009 and many Japanese awards. \par
\end{document}